\documentclass[acmsmall]{acmart}

\usepackage{booktabs} 
\usepackage{flushend}
\usepackage[ruled]{algorithm2e} 
\usepackage{multirow}
\usepackage{subfigure}
\usepackage{tcolorbox,threeparttablex,longtable,afterpage}
\usepackage{latexsym}
\usepackage{amsmath}
\usepackage{enumitem}
\usepackage{color}
\usepackage{xcolor,colortbl}
\usepackage{tcolorbox,tabularx}
\usepackage{makecell}
\newcommand{\tabincell}[2]{\begin{tabular}{@{}#1@{}}#2\end{tabular}}

\usepackage{tikz}
\newcommand*{\circled}[1]{\lower.7ex\hbox{\tikz\draw (0pt, 0pt)%
    circle (.5em) node {\makebox[1em][c]{\small #1}};}}

\newcommand{\num}{100\xspace} 

\AtBeginDocument{%
  \providecommand\BibTeX{{%
    \normalfont B\kern-0.5em{\scshape i\kern-0.25em b}\kern-0.8em\TeX}}}

\setcopyright{acmcopyright}
\acmYear{2023} \acmVolume{1} \acmNumber{1} \acmArticle{1} \acmMonth{7} \acmPrice{15.00}

\begin{document}

\title{Fairness Testing: A Comprehensive Survey and Analysis of Trends}


\author{Zhenpeng Chen}
\affiliation{%
  \institution{University College London}
  \city{London}
  \country{United Kingdom}
}

\author{Jie M. Zhang}\authornote{Corresponding author: Jie M. Zhang (jie.zhang@kcl.ac.uk).}
\affiliation{%
  \institution{King's College London}
  \city{London}
  \country{United Kingdom}
}

\author{Max Hort}
\affiliation{%
  \institution{Simula Research Laboratory}
  \city{Oslo}
  \country{Norway}
}

\author{Mark Harman}
\affiliation{%
  \institution{University College London}
  \city{London}
  \country{United Kingdom}
}

\author{Federica Sarro}
\affiliation{%
  \institution{University College London}
  \city{London}
  \country{United Kingdom}
}

\renewcommand{\shortauthors}{Z. Chen et al.}

\begin{abstract}
Unfair behaviors of Machine Learning (ML) software have garnered increasing attention and concern among software engineers. To tackle this issue, extensive research has been dedicated to conducting fairness testing of ML software, and this paper offers a comprehensive survey of existing studies in this field. 
We collect \num papers and organize them based on the testing workflow (i.e., how to test) and testing components (i.e., what to test). Furthermore, we analyze the research focus, trends, and promising directions in the realm of fairness testing. We also identify widely-adopted datasets and open-source tools for fairness testing.
\end{abstract}


\authorsaddresses{Zhenpeng Chen, Mark Harman, and Federica Sarro are with the Department of Computer Science, University College London, London, United Kingdom. Emails: \{zp.chen, mark.harman, f.sarro\}@ucl.ac.uk. Jie M. Zhang is with the Department of Informatics, King's College London, London, United Kingdom. E-mail: jie.zhang@kcl.ac.uk. Max Hort is with the Simula Research Laboratory, Oslo, Norway. E-mail: maxh@simula.no.}

\maketitle

\section{Introduction}\label{intro}
Machine Learning (ML)-enabled software, commonly referred to as ML software, has gained widespread adoption in critical areas of society, including hiring \cite{chan2018hiring}, credit assessment \cite{bono2021algorithmic}, and criminal justice \cite{berk2021fairness}. However, the use of such software has also led to instances of unfair decision-making, particularly when sensitive attributes such as sex, race, age, and occupation are involved~\cite{sigsoftBrunM18}. For instance, an ML-enabled recidivism assessment system employed by US courts was found to incorrectly label black defendants as higher-risk individuals compared to white defendants \cite{faircase}.

The unfair behaviors exhibited by ML software can have profound ethical implications, resulting in  unacceptable outcomes, particularly by disadvantaging minority groups and protected categories. Consequently, there is a growing concern and heightened awareness within the research community regarding the issue of unfairness and its impact. 

The exploration of fairness issues gained significant momentum in the late 1960s \cite{fatHutchinsonM19}, when psychometricians began investigating the fairness of educational tests \cite{cleary1966test,cleary1968test}. In the 1980s, researchers delved into the social impact of technology \cite{langdon1986artifacts}. Subsequently, the study of fairness expanded to include the domains of ML \cite{kddPedreschiRT08} and Software Engineering (SE) \cite{FinkelsteinHMRZ08}, beginning around 2008, in response to the rapid adoption of ML in software applications supporting decision-making processes.

From the SE perspective, fairness is a non-functional software property that should be treated as a first-class entity throughout the entire SE process \cite{sigsoftBrunM18,icseAydemirD18a}. Ahmad et al. \cite{AhmadBA0G21} emphasized the increasing importance of fairness as a requirement that should be considered during the \emph{requirements engineering} phase of ML software development. Alidoosti \cite{alidoosti2021ethics} argued for the inclusion of fairness considerations in the \emph{design} process of software architecture. Albarghouthi et al. \cite{AlbarghouthiDDN17} framed fairness as correctness properties for ML program \emph{verification}. Zhang et al. \cite{tseZhangHML22} described fairness as a significant \emph{testing} property for ML software. Additionally, Zhang et al. \cite{JiangZhangcorr} viewed fairness as the objective of \emph{repairing} ML software.

In this paper, we also approach fairness from the SE perspective and define imperfections in software systems that result in a misalignment between desired fairness conditions and actual outcomes as \emph{fairness bugs}. Our focus is on \emph{fairness testing} of ML software, which aims to uncover fairness bugs through code execution. Fairness testing represents an important aspect of software fairness research and is closely intertwined with other activities in the SE process. It verifies whether software systems meet fairness requirements, exposes fairness bugs introduced during software implementation, and guides software repair efforts to address fairness issues, among other related SE activities.

Compared to traditional software testing, fairness testing presents distinct challenges. For example, in traditional software testing, a test oracle typically relies on the output of a single input. In fairness testing, the oracle problem is more challenging because inputs and outputs from different demographic groups need to be considered simultaneously. Additionally, there are diverse fairness definitions, some of which may even conflict with each other or be mathematically impossible to satisfy concurrently \cite{conficseVermaR18,castelnovo2022clarification}. Since different definitions can require different test oracles, designing fairness testing techniques to accommodate this multitude of definitions poses a significant problem as well.

\begin{figure}[t]
    \centering
      \includegraphics[width=0.5\linewidth]{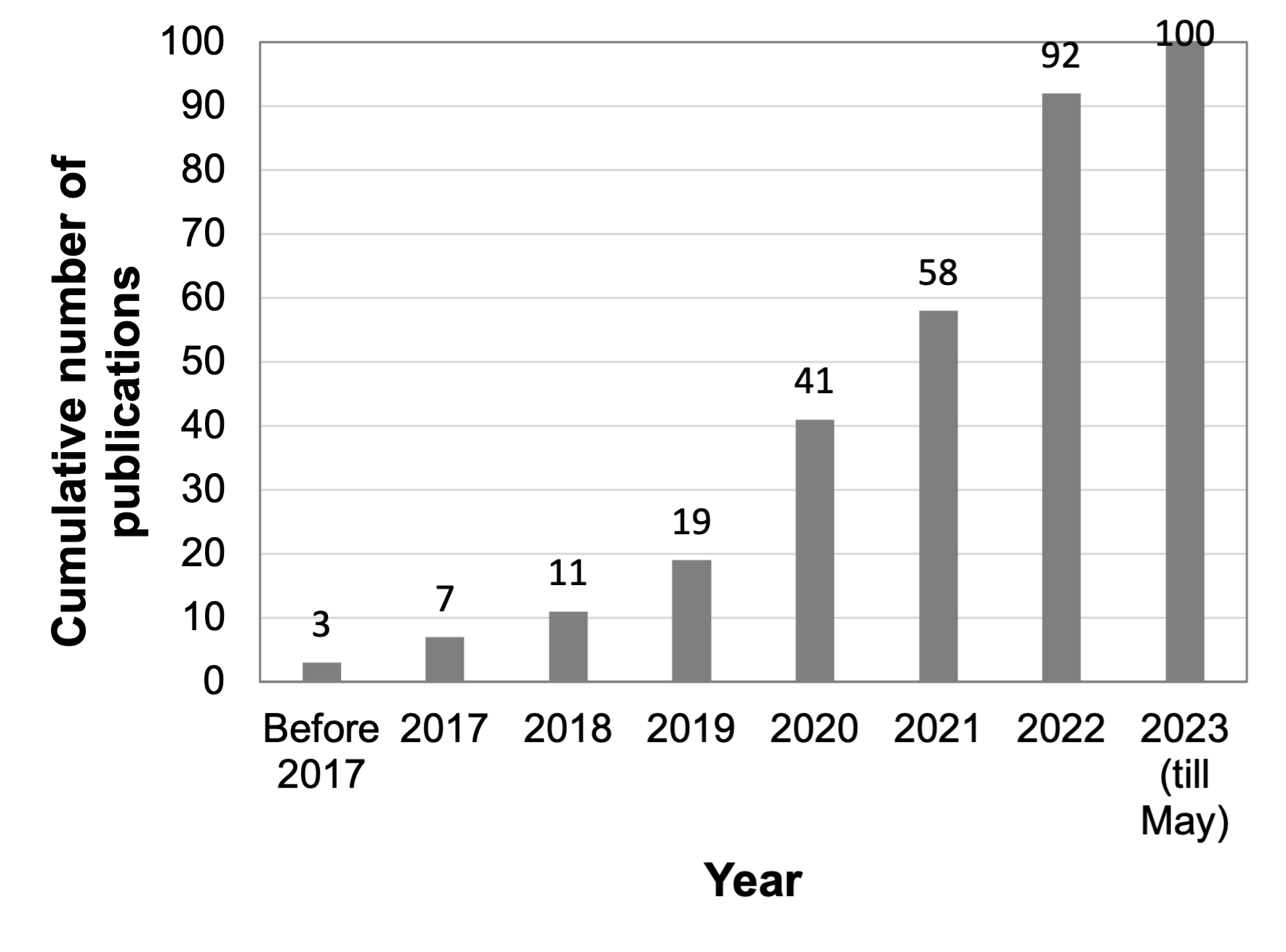}
  \caption{Cumulative number of publications on fairness testing.}
  \label{fig:trend} 
\end{figure}

The significance of fairness testing and its associated challenges has led to a notable increase in research efforts in this field. Figure \ref{fig:trend} illustrates the cumulative number of publications on fairness testing until 2023, revealing a growing interest and emphasizing the relevance of this survey. Notably, 89\% of fairness testing publications have emerged since 2019, indicating the emergence of this new domain of software testing.

This paper offers a comprehensive survey of fairness testing in ML software. The collected papers are sourced from various venues including SE, artificial intelligence, computer security, and human-computer interaction. We categorize these papers based on two key aspects: the fairness testing workflow (i.e., how to test) and fairness testing components (i.e., what to test). Furthermore, we conduct an analysis of research trends and identify potential research opportunities for the fairness testing community. Additionally, we provide an overview of publicly accessible datasets and open-source tools available for fairness testing.

Previous surveys have explored various aspects of fairness in ML and related fields. Mehrabi et al. \cite{csurMehrabiMSLG21} and Pessach and Shmueli \cite{fairnessreview22} surveyed fairness research on ML algorithms, while Hort et al. \cite{maxfairnesssurvey} focused on bias\footnote{The terms ``bias'' and ``unfairness'' are often used interchangeably in the literature, as they both denote deviations from ``fairness'' \cite{csurMehrabiMSLG21}.} mitigation methods for ML classifiers. Sun et al. \cite{sunetal2019mitigating}, Berk et al. \cite{berk2021fairness}, and Pitoura et al. \cite{vldbFairness} surveyed techniques for improving fairness in specific ML tasks, such as natural language processing, criminal justice risk assessment, and ranking. Tushev et al.~\cite{9669016} surveyed software design strategies for fairness in digital sharing economy applications. Hutchinson and Mitchell \cite{fatHutchinsonM19} provided a historical perspective on fairness assessment tests across disciplines, including education and hiring. Zhang et al. \cite{tseZhangHML22} conducted a broader survey on ML testing, considering fairness as one of several testing properties. A recent systematic literature review~\cite{corrabs220508809} focused on software fairness, covering only 20 fairness testing papers. In contrast, our survey specifically concentrates on fairness testing of ML software. To the best of our knowledge, this is the first comprehensive survey specifically dedicated to the literature on fairness testing.

To summarize, this work makes the following contributions:

\begin{itemize}
\item It provides a comprehensive survey of \num fairness testing papers, encompassing diverse research communities.
\item It defines fairness bug and fairness testing, and provides an overview of the testing workflow and testing components related to fairness testing. 
\item It compiles a summary of public datasets and open-source tools for fairness testing, providing a navigation for researchers and practitioners interested in the field.
\item It analyzes research trends and identifies promising research opportunities in fairness testing, aiming to foster further advancements in this area.
\end{itemize}

\begin{figure}[t]
    \centering
       \includegraphics[width=0.8\linewidth]{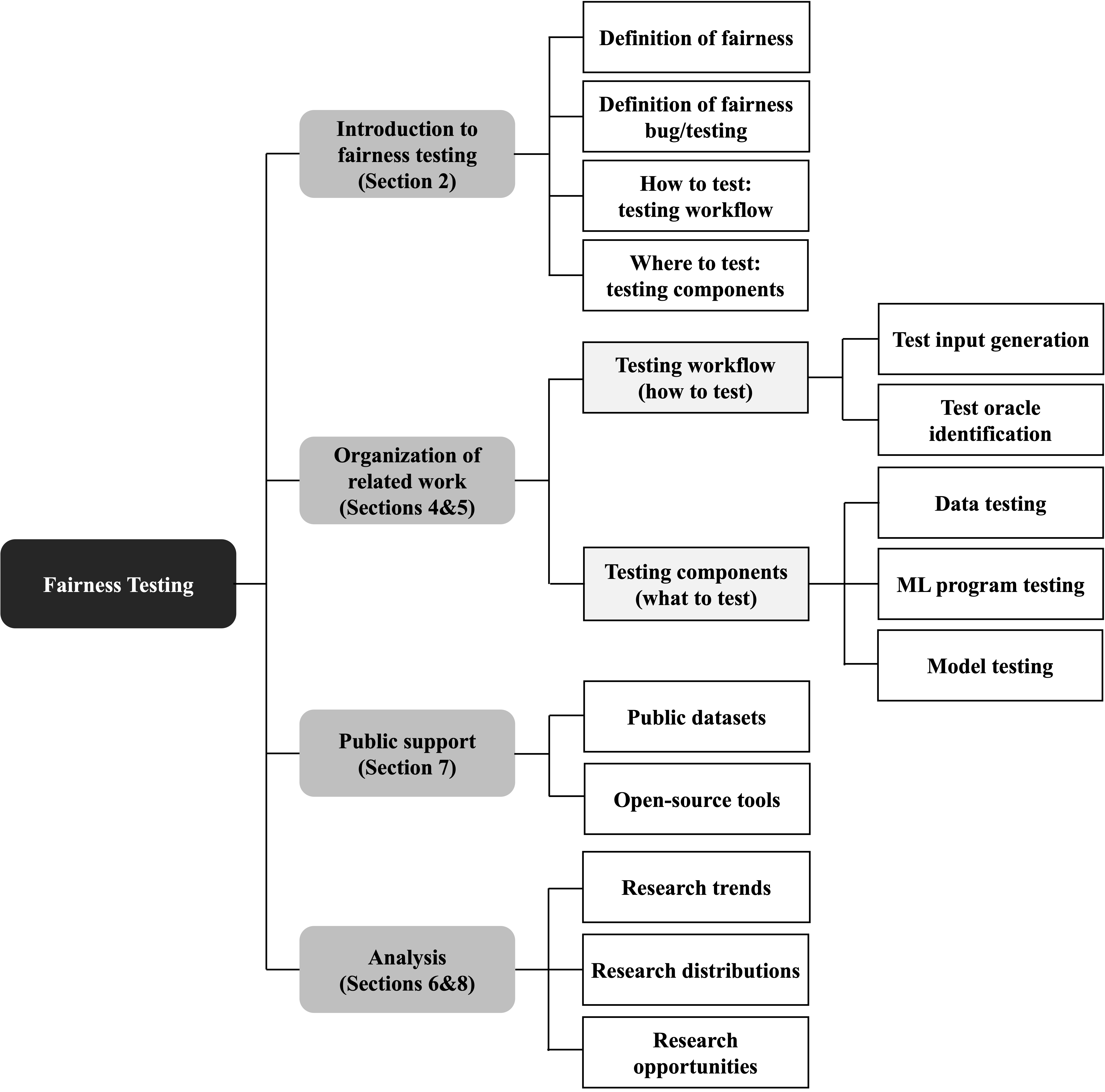}
  \caption{Structure of this paper.}
  \label{fig:struc} 
\end{figure}

The structure of the paper is illustrated in Figure \ref{fig:struc}, and the detailed survey methodology is presented in Section \ref{methodology}.

\section{Preliminaries}\label{preliminaries}
In this section, we begin by presenting widely-adopted fairness definitions. Subsequently, we offer a definition of fairness bug and fairness testing from the perspective of SE. We further outline the testing workflow and testing components of fairness testing. 

\subsection{Definition of Fairness}\label{fdefinition} 
The definition of fairness plays a crucial role in establishing the fairness conditions that software systems are expected to meet. Over the years, researchers and practitioners have proposed and explored various fairness definitions \cite{hellman2020measuring,mitchell2021algorithmic,castelnovo2022clarification}. This section aims to present the fairness definitions that have received widespread adoption in the literature \cite{tseZhangHML22}, as listed in Table \ref{fairness_metric}. 

These definitions primarily fall into two categories: \emph{individual fairness} and \emph{group fairness} \cite{csurMehrabiMSLG21}. Individual fairness requires that software should produce similar predictive outcomes for similar individuals, while group fairness requires software to treat different demographic groups in a similar manner. Fairness assessment in the context of ML software often relies on \emph{sensitive attributes}, also known as \emph{protected attributes}, which represent characteristics that require protection against unfairness, such as sex, race, age, and physical ability. By considering sensitive attributes, the population can be categorized into privileged groups and unprivileged groups.

\begin{table}[!tp]
\small 
\centering
\caption{Widely-adopted fairness definitions.}
\label{fairness_metric}
\begin{tabular}{ll}
\hline
Name &  Fairness type \\
\hline
Fairness through unawareness \cite{grgic2016case} &  Individual fairness \\
Fairness through awareness \cite{innovaDworkHPRZ12} &  Individual fairness\\
Counterfactual fairness \cite{nipsKusnerLRS17} & Individual fairness \\
Causal fairness \cite{sigsoftGalhotraBM17} & Individual fairness\\
Statistical parity \cite{barocas2016big} & Group fairness \\
Equalized odds \cite{nipsHardtPNS16} & Group fairness \\
Equal opportunity \cite{nipsHardtPNS16} & Group fairness \\
\hline
\end{tabular}
\end{table}

To facilitate the formalization of fairness, we introduce the necessary notations:

\begin{itemize}

\item[-] $X$: Denotes the input feature vector of ML software, excluding sensitive attributes. $X^{(i)}$ denotes the feature vector for individual $i$.

\item[-] $A$: Denotes sensitive attributes, where $A^{(i)}$ denotes the sensitive attributes for individual $i$.

\item[-] $Y$: Denotes the actual outcome of ML software.

\item[-] $\hat{Y}$: Denotes the predicted outcome, with $\hat{Y}(X^{(i)}, A^{(i)})$ indicating the predicted outcome for individual $i$.


\item[-] $P(\cdot)$: Denotes the probability function.
\end{itemize}

For $Y$ and $\hat{Y}$, we use $F$ to denote a favorable outcome and $UF$ to denote an unfavorable outcome.
Given that ML inputs of diverse data types can be encoded as vectors, these definitions of fairness are applicable to a wide range of data types.

\begin{table}[!tp]
\scriptsize
\centering
\caption{Advantages and disadvantages of different individual fairness definitions.}
\label{inf_addis}
\begin{tabularx}{\linewidth}{p{1.8cm}p{5.5cm}p{5.5cm}}
\hline
Name &  Advantages & Disadvantages\\
\hline
Fairness through unawareness & Straightforward solution by avoiding explicit use of sensitive attributes. & Ignores correlations between non-sensitive features and sensitive attributes.\\
\hline
Fairness through awareness & (1) Considers both the similarity of individuals and the similarity of outcome distributions. (2) Flexible similarity definition for different scenarios. & (1) Choice of distance metrics can impact results and may require fine-tuning. (2) Sensitivity to the definition of similarity, which can vary across scenarios.\\
\hline
Counterfactual fairness & (1) Incorporates causal reasoning to identify discriminatory effects. (2) Considers an impact of changes in sensitive attributes on both non-sensitive features and predictions. & (1) Requires knowledge of causal relationships between non-sensitive features and sensitive attributes. (2) Practical implementation may be challenging, especially when causal relationships are complex.\\
\hline
Causal fairness & (1) Incorporates causal reasoning to identify discriminatory effects. (2) Easy to implement compared to counterfactual fairness. & Ignores potential relationships between non-sensitive features and sensitive attributes.\\
\hline
\end{tabularx}
\end{table}

\begin{table}[!tp]
\scriptsize
\centering
\caption{Advantages and disadvantages of different group fairness definitions.}
\label{groupf_addis}
\begin{tabularx}{\linewidth}{p{1.8cm}p{5.5cm}p{5.5cm}}
\hline
Name &  Advantages & Disadvantages\\
\hline
Statistical parity & (1) Easy to understand and implement. (2) Promotes fairness at a high level, focusing on overall outcome balance. & (1) May not address nuanced biases in specific performance metrics. (2) Ignores variations in predictive performance between groups.\\
\hline
Equalized odds & (1) Considers both positive and negative predictive performance. (2) Addresses potential disparities in error rates between groups. & (1) Stricter conditions may be challenging to satisfy in practice. (2) May be sensitive to class imbalances and prevalence differences.\\
\hline
Equal opportunity & (1) Focuses specifically on fairness in positive predictions. (2) Emphasizes equal opportunities for positive outcomes. & (1) Does not consider false positive rates, potentially overlooking negative consequences. (2) Similar to equalized odds, may face challenges in practical implementation.\\
\hline
\end{tabularx}
\end{table}


\subsubsection{Individual Fairness}
We first introduce four widely-adopted definitions of individual fairness.


\textbf{Fairness through unawareness} \cite{grgic2016case} assumes that a software system can achieve fair outcomes by refraining from explicitly using sensitive attributes in the decision-making process. By excluding these attributes, the system cannot rely on them and is thus expected to produce the same outcome for individuals with identical non-sensitive features. 
Formally, fairness through awareness assumes that, for individuals $i$ and $j$, if $X^{(i)} = X^{(j)}$, then $\hat{Y}(X^{(i)}) = \hat{Y}(X^{(j)})$.

\textbf{Fairness through awareness} \cite{innovaDworkHPRZ12} requires a software system to produce similar outcomes for similar individuals. To achieve this, two distance metrics are employed: $d(\cdot, \cdot)$ measures the similarity of individuals for a specific task, and $D(\cdot, \cdot)$ measures the similarity of probability distributions over outcomes. Formally, fairness through awareness dictates that $D(\hat{Y}(X^{(i)}, A^{(i)}), \hat{Y}(X^{(j)}, A^{(j)})) \leq d((X^{(i)}, A^{(i)}), (X^{(j)}, A^{(j)}))$. In other words, the distributions over predicted outcomes for two individuals should be indistinguishable within their measured similarity/distance. Thus, if two individuals are similar, they should receive similar predicted outcomes.


\textbf{Counterfactual fairness} \cite{nipsKusnerLRS17} states that an individual's prediction should remain the same in the real world as well as in a counterfactual world where the individual belongs to a different demographic group (i.e., the sensitive attribute is different). In practical applications, the input $X$ may contain features that have a causal relationship with the sensitive attribute $A$. Therefore, when the attribute $A=a$ is changed to the counterfactual value $A=a'$, the input $X=x$ will also be transformed to $X=x'$, where features causally linked to the sensitive attribute will be altered accordingly. Formally, under any context $X =x$ and $A=a$: $P(\hat{Y}(x, a)=y|X=x, A=a) = P(\hat{Y}(x', a')=y|X=x, A=a)$, for all $y$ and for any value $a'$ attainable by $A$.

\textbf{Causal fairness} \cite{sigsoftGalhotraBM17} captures the causal relationship between sensitive attributes and outcomes. It involves running a software system on an input, modifying the sensitive attribute(s), and checking whether this modification leads to a change in the output. Formally, causal fairness requires that for any individual $i$, there should be no other instance $j$ that simultaneously satisfies the following conditions: (1) $X^{(i)} = X^{(j)}$; (2) $A^{(i)} \neq A^{(j)}$; (3) $\hat{Y}(X^{(i)}, A^{(i)}) \neq \hat{Y}(X^{(j)}, A^{(j)})$.


A notable distinction between fairness through unawareness and other fairness definitions lies in its deliberate exclusion of sensitive attributes from the decision-making process. In contrast, fairness through awareness incorporates sensitive attribute information depending on the definition of individual similarity; counterfactual fairness and causal fairness make use of sensitive attributes to comprehend their relationship with outcomes.


Next, we discuss the difference among fairness through awareness, counterfactual fairness, and causal fairness. Fairness through awareness is a more general concept compared to counterfactual fairness and causal fairness. The alignment between fairness through awareness and counterfactual fairness emerges when the similarity metric is crafted by contemplating the likeness of samples in counterfactual worlds \cite{nipsKusnerLRS17}. On the other hand, when the similarity metric is grounded in the similarities of non-sensitive attributes, fairness through awareness can be likened to causal fairness~\cite{conficseVermaR18}. While counterfactual fairness and causal fairness share similarities in their approach to individual fairness—both adopting a causal standpoint that evaluates fairness by manipulating sensitive attribute information—they diverge in certain aspects. 
Specifically, causal fairness simplifies this perspective by modifying only sensitive attribute information without considering the causal relationships among sensitive attributes and other features. In contrast, counterfactual fairness changes both sensitive attribute information and non-sensitive features that are causally linked to sensitive attributes.

Table \ref{inf_addis} summarizes advantages and disadvantages of different individual fairness definitions.



\subsubsection{Group Fairness}
We introduce three widely-adopted definitions of group fairness. To better illustrate the distinctions between these definitions, we use a credit score software system as an example, with sex as the protected attribute of applicants.

\textbf{Statistical parity} \cite{barocas2016big}, also known as \textbf{demographic parity}, requires the probability of a favorable outcome to be the same among different demographic groups. In other words, a software system satisfies statistical parity if, for any two possible values (denoted as $a$ and $a'$) of the sensitive attribute set, the probability of obtaining favorable outcomes is the same: $P[\hat{Y}=F|A=a]=P[\hat{Y}=F|A=a']$.
For instance, the credit score system should assign a good score to male and female applicants in equal proportions.

\textbf{Equalized odds} \cite{nipsHardtPNS16} requires that the privileged and the unprivileged groups have equal true positive rates and equal false positive rates.
In other words, the prediction is independent of sensitive attributes when the target label $Y$ is fixed. For any two possible values (denoted as $a$ and $a'$) of the sensitive attribute set, $P[\hat{Y}=F|A=a,Y=y]=P[\hat{Y}=F|A=a',Y=y]$ for $y\in\{F,UF\}$.
In our example, the probability of correctly assigning a good predicted credit score to an applicant with an actual good credit score, and the probability of incorrectly assigning a good predicted credit score to an applicant with an actual bad credit score, should be equal for male and female applicants.

\textbf{Equal opportunity} \cite{nipsHardtPNS16} states that the privileged and the unprivileged groups have equal true positive rates. In other words, the prediction is independent of sensitive attributes when the target label $Y$ is fixed as $F$: for any two possible values (denoted as $a$ and $a'$) of the sensitive attribute set, $P[\hat{Y}=F|A=a,Y=F]=P[\hat{Y}=F|A=a',Y=F]$. In our example, the probability of accurately assigning a good predicted credit score to an applicant with an actual good credit score should be the same for male and female applications.


Table \ref{groupf_addis} summarizes advantages and disadvantages of different group fairness definitions.

More about fairness definitions can be found in recent work on surveying, analyzing, and comparing them. Mitchell et al. \cite{mitchell2021algorithmic} compiled a comprehensive catalog of fairness definitions found in the literature, offering a valuable resource for understanding different perspectives. Verma and Rubin \cite{conficseVermaR18} provided insights into the rationale behind existing fairness definitions and examined their relationships through a detailed case study. Castelnovo et al. \cite{castelnovo2022clarification} delved into the differences, implications, and orthogonal aspects of various fairness definitions, shedding light on their distinct characteristics. Additionally, Mehrabi et al. \cite{csurMehrabiMSLG21} developed a taxonomy of fairness definitions proposed specifically for ML algorithms.

\subsection{Definition of Fairness Bug and Fairness Testing}\label{defftesting}
A software bug is an imperfection in a computer program that causes a discordance between existing and required conditions \cite{5399061}. Based on this definition, prior research \cite{tseZhangHML22} defines ``ML bug'' as any imperfection in an ML item that causes a discordance between the existing and the required conditions, and ``ML testing'' as any activity designed to reveal ML bugs. Aligning with the terminology used in SE, we define fairness bug and fairness testing as follows.

\noindent $\bullet$ \textbf{Definition (Fairness Bug).} \emph{A fairness bug refers to any imperfection in a software system that causes a discordance between the existing and required fairness conditions.}

The required fairness conditions depend on the fairness definition adopted by the developers of the software under test. The imperfections can manifest as unfair predictions that violate individual fairness or discrepancies in outcomes among different demographic groups that surpass a predetermined threshold for group fairness. Previous studies have also referred to such imperfections as fairness defects \cite{sigsoftBrunM18} or fairness issues \cite{FinkelsteinHMRZ08}. In this paper, we adopt the term ``fairness bug'' as a representative of these related concepts, as ``bug'' carries a broader meaning \cite{5399061}. 

The presence of fairness bugs has spurred research efforts towards developing automated testing techniques to detect these bugs, i.e., fairness testing.

\noindent $\bullet$ \textbf{Definition (Fairness Testing)}. \emph{Fairness testing refers to any activity designed to reveal fairness bugs through code execution.} Formally, consider a software system  $S$ under test, a set of inputs $I$, the required fairness condition $C$, and the existing fairness condition $C'$. Fairness testing refers to any activity involving the execution of $I$ on $S$ to identify a discordance between $C$ and $C'$.

This definition categorizes fairness testing as a subset of software testing, specifically excluding manual inspection and formal verification methods.
Since we focus on fairness testing of ML software, we follow the recent ML testing survey \cite{tseZhangHML22} to consider two key aspects of this emerging testing domain: \emph{testing workflow} and \emph{testing components}.

\subsection{Fairness Testing Workflow}\label{ftworkf}
The fairness testing workflow refers to \emph{how} to conduct fairness testing with different testing activities. 
This section outlines its key activities.

\begin{figure*}[t]
    \centering
       \includegraphics[width=1\linewidth]{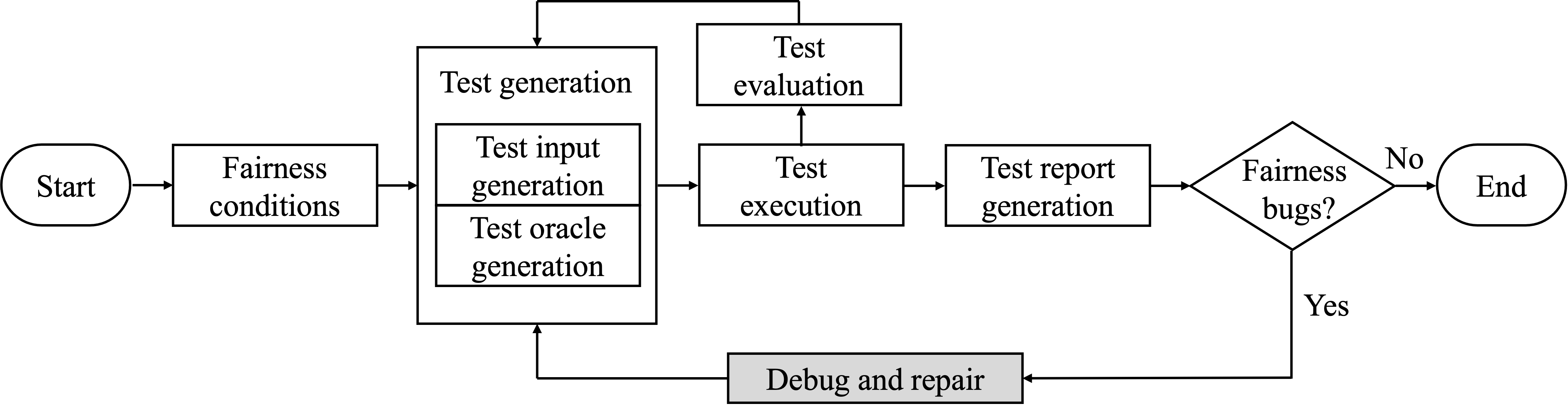}
  \caption{Workflow of fairness testing.}
  \label{fig:workflow} 
\end{figure*}

Figure \ref{fig:workflow} presents an overview of the fairness testing workflow for ML software, which builds upon the established ML testing workflow in previous work \cite{tseZhangHML22}. In this process, software engineers establish and specify the desired fairness conditions (also often referred to as fairness requirements) for the software under test. These fairness conditions  primarily stem from individual fairness or group fairness definitions. Based on the defined fairness conditions, test oracles are identified and created. Test inputs are obtained either through sampling from collected data or generation. The test inputs are then executed on the software under test to determine if any violations of the fairness conditions occur. Engineers assess the adequacy of the tests, evaluating their effectiveness in uncovering fairness bugs. Simultaneously, the results of the test execution provide a bug report that assists engineers in reproducing, locating, and resolving any identified fairness bugs. The fairness testing process can be iterated to ensure that the repaired software aligns with the desired fairness conditions. Upon satisfying the fairness conditions, the software can be deployed for use. Following deployment, engineers can employ runtime monitoring to continually test the software using new real-world inputs, ensuring ongoing adherence to the fairness conditions.

Note that, as in ML testing \cite{tseZhangHML22}, test inputs in fairness testing can take diverse forms based on the testing components of ML software. Common testing components include data, ML programs, and ML models, as described in Section \ref{component_pre}. For testing data, test inputs can be programs capable of detecting data bias \cite{tseZhangHML22}; for testing ML programs and models, test inputs typically involve data instances used to validate ML behaviors \cite{tseZhangHML22}. Unlike traditional software testing, which can rely on a single data instance as the test input, fairness testing of ML programs and models typically uses pairs of data instances (for individual fairness testing) or a set of data instances with diverse demographic information (for group fairness testing).

Section \ref{workflowsection} organizes fairness testing papers based on the testing workflow perspective. The existing fairness testing literature primarily addresses test input generation (Section \ref{input_generation}) and test oracle identification (Section \ref{oracle}), while other testing activities remain open challenges and present research opportunities for the community (discussed in Section \ref{chall}).

\subsection{Fairness Testing Components}\label{component_pre}
Software testing can be conducted for different testable parts within a software system \cite{LaghariD18}. In the context of fairness testing, this section focuses on introducing the testable components within ML software.

\begin{figure}[t]
    \centering
       \includegraphics[width=0.4\linewidth]{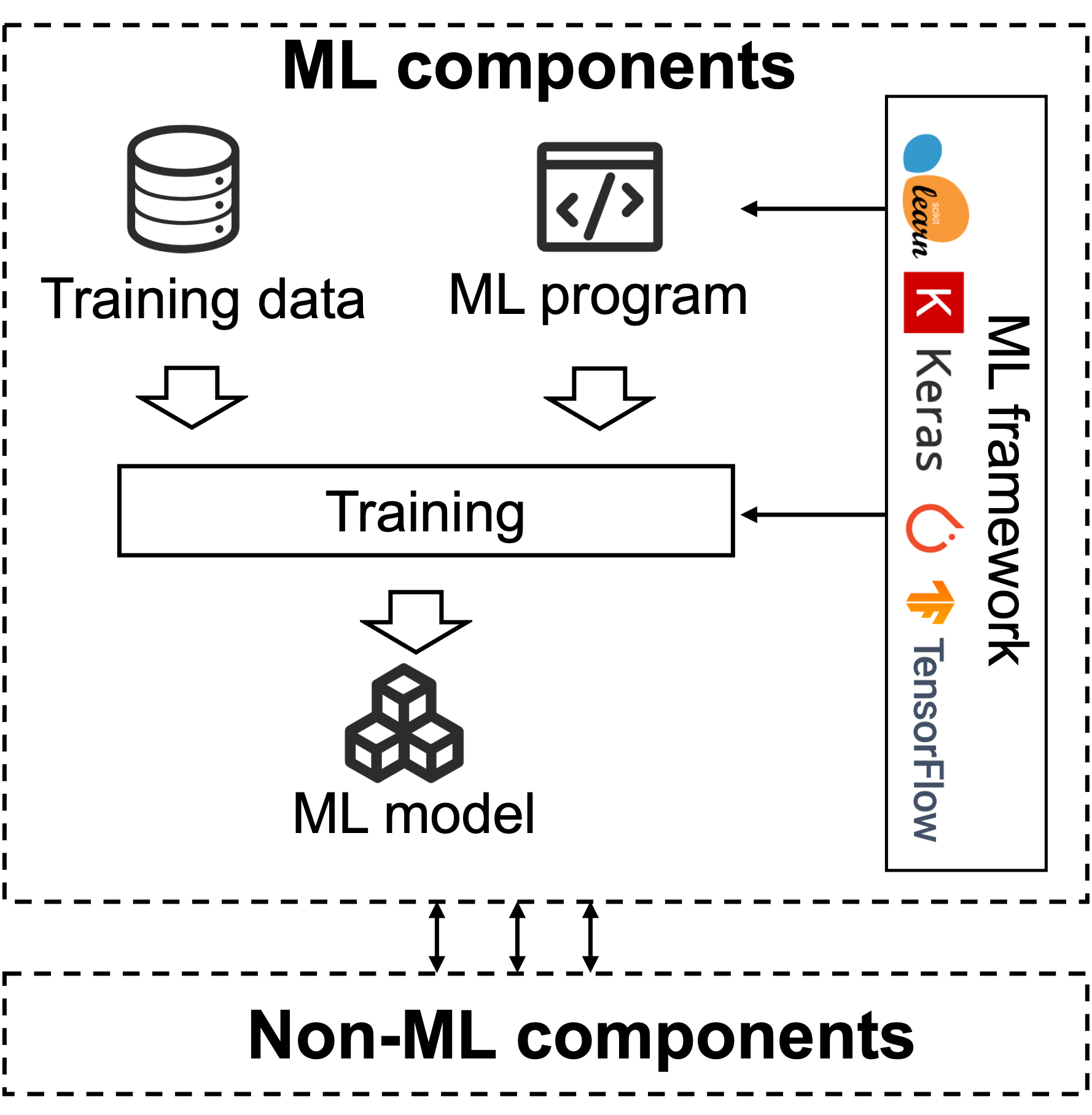}
  \caption{Components to test in ML software.}
  \label{fig:compo} 
\end{figure}

We denote the entire ML software used by end-users as $S$, where $S$ encompasses both ML components and non-ML components~\cite{icseLuZXWDS22, icseNaharZLK22}, as illustrated in Figure~\ref{fig:compo}. In the ML components, an ML model $M$ is trained using large-scale data $D$, comprising sensitive attributes $A$, remaining feature vectors 
$X$, and data labels $Y$. An ML program $G$ is implemented based on ML frameworks $F$ (e.g., scikit-learn \cite{pedregosa2011scikit}, TensorFlow \cite{osdiAbadiBCCDDDGIIK16}, and Keras \cite{gulli2017deep}) \cite{tseZhangHML22, sigsoftChenCLW0L20}, and it specifies the structure of the desired ML model and the process of obtaining it through the training data \cite{sigsoftChenCLW0L20}. To translate ML components into a functional software system, various non-ML components denoted as $C$ are necessary, such as data storage, user interfaces, and monitoring infrastructures. All these components are interconnected and work together to achieve the software’s objective.

In Section \ref{defftesting}, we define fairness testing as any activity aimed at uncovering fairness bugs in the ML software $S$. According to this definition, fairness testing for $S$ should encompass testing activities on each component, including the training data $D$, ML program $G$, ML frameworks $F$, ML model $M$, and non-ML components $C$. This layered approach to fairness testing addresses diverse aspects of the ML software $S$, facilitating a comprehensive evaluation that identifies and rectifies biases at various stages. The collaboration of these testing components collectively fulfills the mission of fairness testing in the ML software.


\textbf{Data Testing.} ML follows the data-driven programming paradigm. Training data $D$ determines the decision logic of ML models $M$ to a large extent \cite{tseZhangHML22}. Therefore, training data is considered as an important component to test in the ML testing literature \cite{tseZhangHML22,mlsysBreckP0WZ19}. Since bias in training data is demonstrated to be a main root cause of ML software unfairness \cite{sigsoftChakrabortyMM21}, training data is also an important component for fairness testing. 
In the fairness literature, data testing aims to identify bias within the training data $D$. As elucidated in Section \ref{ftworkf}, data testing involves the use of programs as test inputs, designed to specifically detect three types of data bias: feature bias, label bias, and selection bias. Feature bias detection focuses on identifying whether the features $X$ of training data contain bias \cite{icse22fair2}. Label bias detection aims to identify whether factors unrelated to label determination influence the label generation process, resulting in biased data labels $Y$ \cite{sigsoftChakrabortyM0M20}. Selection bias detection targets the identification of unexpected correlations between sensitive attributes $A$ and data labels $Y$ within the distribution of training data \cite{nipsWickpT19}.


\textbf{ML Program Testing.} An ML program $G$ encodes the process by which an ML model $M$ is obtained based on the training data. It has been an important fairness testing component of ML software~\cite{tseZhangHML22}. A fairness bug may arise as improper data processing \cite{biswas2021fair}, training algorithm selection \cite{sigsoftHortZSH21}, and hyper-parameter settings (i.e., configuration options that control the learning process) \cite{icse22fair1} in ML programs. In the fairness literature, ML program testing aims to uncover issues within the implementation of the ML program $G$ that contribute to the unfairness of the ML software $S$.

\textbf{Framework Testing.} ML frameworks $F$ (also called ML libraries) implement ML algorithms internally and provide high-level Application Programming Interfaces (APIs) for developers to build ML models without knowledge of the inner working of these algorithms. The ML testing literature \cite{icsePhamLQT19,sigsoftWangYCLZ20,NejadgholiY19,icseWangLQP022,MoussaMLtools} has considered ML frameworks as an important testing component and detected bugs inside ML frameworks that lead to accuracy problems in the final ML model $M$. Nevertheless, existing framework testing studies are primarily related to ML performance (e.g., accuracy). In contrast, framework testing for fairness aims to detect issues inside ML frameworks $F$ that result in unfairness of the ML software $S$. To the best of our knowledge, to date, there has been no framework testing work for detecting fairness bugs.

\textbf{Model Testing:} Most fairness testing techniques are model-centric \cite{sigsoftGalhotraBM17,icseMingFan,kbseUdeshiAC18,sigsoftAggarwalLNDS19,trustcomXieW20,icseZhangW0D0WDD20,isstaZhangZZ21,icse22fairNEURONFAIR}. They consider the ML model $M$ as the testing component and aim to reveal fairness bugs based on the input-output behaviors of $M$. Fairness testing of ML models can be performed in a white, black, or gray-box manner \cite{icse22fair1}. Black-box testing is a technique of testing without having any knowledge of the internal working of ML software (e.g., code and data); white-box testing tests an ML software system taking into consideration its internal working; gray-box testing is to test with limited knowledge of the internal working of the software under test \cite{khan2012comparative}. 

\textbf{Non-ML component testing:} From the SE perspective, ML software is beyond the ML models and also includes non-ML components $C$ \cite{DBLPjournalscorrabs2111}. These non-ML components may also affect the fairness of ML software. For instance, data storage practices that result in the exclusion or underrepresentation of specific demographic groups can lead to biased training data, subsequently resulting in biased ML models. Similarly, biased user interfaces may inadvertently exhibit favoritism or bias against particular user groups. Testing non-ML components for fairness seeks to identify any unfairness within the ML software $S$ attributable to the non-ML components $C$. However, to the best of our knowledge, there has been no non-ML component testing work in the fairness testing literature.

Section \ref{testcomponent} organizes the related papers from the testing component perspective. The existing literature primarily focuses on data testing (Section \ref{datatesting}), ML program testing (Section \ref{programtesting}), and model testing (Section \ref{modeltesting}).

\section{Survey Methodology}\label{methodology}
This section introduces the scope and the paper collection process of our survey.

\subsection{Survey Scope}
We aim to define, collect, and curate the disparate literature, arguing and demonstrating that there does, indeed, exist a coherent area of research in the field that can be termed ``fairness testing''.

We apply the following inclusion criteria when collecting papers. The papers that satisfy any of these criteria are included in this survey.

\begin{itemize}[leftmargin=*]
  \item The paper introduces the general idea or one of the related aspects of fairness testing of ML software.
    \item The paper presents an approach, study, framework, or tool that targets at fairness testing of ML software.
\end{itemize}

We do not include papers about the issues of fairness in network systems and hardware systems. Moreover, we filter out papers that are about fairness definitions but do not consider them in the context of testing.
We also do not include papers about gender diversity/inclusion and cognitive bias in software development, because our survey focuses on SE product fairness, not SE process fairness.

\subsection{Paper Collection}
We first collected papers by keyword searching on the DBLP publication database \cite{dblplink}, which covers arXiv (a widely-used open-access archive), more than 1,800 journals, and 5,800 academic conferences and workshops in computer science \cite{dblpstatislink}. DBLP has been extensively used in previous surveys in SE \cite{tseZhangHML22,csurChenPPXZHZ20,jcstZhangTJLPY18,Mathewtse,garousi2016highly}, and a recent ML testing survey \cite{tseZhangHML22} demonstrates that papers collected from other popular publication databases are a subset of those collected from DBLP. 

We developed the search keywords through an iterative trial-and-error procedure \cite{tosemLinCSBNL22} conducted by the first two authors, with input and discussion from the other authors. Initially, we started with a general search string ``fairness testing'' to gather initial relevant papers. Subsequently, we carefully examined the titles, abstracts, and keywords of these papers to identify additional keywords and phrases. Through brainstorming sessions, we expanded and refined the list of search strings by incorporating related terms, synonyms, and variations. The iterative process allowed us to continuously improve the search keyword list based on the outcomes of the searches, ensuring that it accurately captured the relevant literature on fairness testing.

The final keywords used for searching included \emph{(``\textbf{fair}'' OR  ``\textbf{bias}'' OR ``\textbf{discrimination}'') AND (``\textbf{software}'' OR ``\textbf{learning}'' OR ``\textbf{bug}'' OR ``\textbf{defect}'' OR ``\textbf{fault}'' OR ``\textbf{algorithm}'' OR ``\textbf{test}'' OR ``\textbf{detect}'' OR ``\textbf{evaluat}'' OR ``\textbf{discover}'' OR ``\textbf{identify}'' OR ``\textbf{find}'' OR ``\textbf{uncover}'' OR ``\textbf{reveal}'' OR ``\textbf{recogniz}'' OR ``\textbf{unveil}'')}.
As a result, we conducted a total of $3\times16=48$ searches on DBLP on March 12, 2023, and obtained 7,674 hits. Then, the first two authors manually inspected each hit paper to check whether it was in the scope of our survey, and selected 67 relevant papers.

The fairness testing community is diverse, with publications appearing in various venues and adopting different terminologies. To capture papers that might have been overlooked by our keywords and ensure a comprehensive coverage of the field, we further employed a snowballing approach. This process, conducted in April and May 2023, aimed to identify transitively dependent papers and expand our paper collection. Both backward and forward snowballing approaches~\cite{esemJalaliW12} were employed.  In \emph{backward snowballing}, we examined the references in each collected paper and identified those lying in our scope; in \emph{forward snowballing}, we used Google Scholar to identify papers of our interest from those that cited the collected papers. 
We iteratively repeated the snowballing process until we reached a fixed point, where no new relevant papers were identified. Through this process, we were able to retrieve an additional 25 papers, contributing to a more comprehensive coverage of the fairness testing literature.

To ensure that our survey is comprehensive and accurate, we also contacted the authors of the papers that we collected via keyword searching and snowballing.
We provided them with our paper and asked them to check whether our description of their work was correct. This interaction allowed us to refine our understanding of their contributions and make necessary revisions to our descriptions. Furthermore, during these communications, authors directed our attention to 15 additional papers that we had not initially included in our collection. Among the suggested papers, 8 met our inclusion criteria and were deemed relevant to our survey. These 8 papers were subsequently added to our repository, further enhancing the coverage of the fairness testing literature in our survey.

Table \ref{collection_info} shows the statistics of the paper collection process. In summary, we consider 67 + 25 + 8 = \num papers for this survey.

\begin{table}[!tp]
\small
\centering
\caption{Statistics of the collected papers.}
\label{collection_info}
\begin{tabular}{lr}
\hline
Keyword & Hits\\
\hline
fair $|$ bias $|$ discrimination + software & 236\\
fair $|$ bias $|$ discrimination + learning & 2,399\\
fair $|$ bias $|$ discrimination + bug & 20 \\
fair $|$ bias $|$ discrimination + defect & 41 \\
fair $|$ bias $|$ discrimination + fault & 94 \\
fair $|$ bias $|$ discrimination + algorithm & 2,121\\
fair $|$ bias $|$ discrimination + test & 409 \\
fair $|$ bias $|$ discrimination + detect & 1,013 \\
fair $|$ bias $|$ discrimination + evaluat & 838\\
fair $|$ bias $|$ discrimination +  discover & 172 \\
fair $|$ bias $|$ discrimination + identify & 119\\
fair $|$ bias $|$ discrimination + find & 96\\
fair $|$ bias $|$ discrimination + uncover & 33 \\
fair $|$ bias $|$ discrimination +  reveal& 60 \\
fair $|$ bias $|$ discrimination +  recogniz& 12\\
fair $|$ bias $|$ discrimination + unveil & 11 \\
\hline
After manual inspection & 67 \\
After snowballing & 92 \\
After collecting author feedback & \num \\
\hline
\end{tabular}
\end{table}

\subsection{Paper Analysis}
To ensure a rigorous analysis of the collected papers and to enhance the quality and accuracy of the survey, we employed thematic synthesis, a well-established method in SE literature reviews~\cite{icseHuangZZBY18}. This method allowed us to systematically organize and analyze the papers in a structured manner.

The analysis process was led by the first two authors, who extensively read and examined the full text of each paper. Their objective was to develop a comprehensive understanding of the testing workflow and test components described in the papers. Through meticulous manual analysis, they extracted relevant information, identified common patterns, and discerned major themes that emerged across the collection of papers.

Using these identified themes as a framework, the authors categorized and organized the related papers under different thematic headings. This approach facilitated a coherent and structured representation of the survey's content, enabling readers to easily navigate and comprehend the key insights and findings from the analyzed papers.

During the analysis, in instances where disagreements arose, the first two authors held discussion meetings involving other co-authors. These meetings served as a platform to address conflicts and reach a consensus on the extracted data and the placement of papers within the identified themes. The co-authors, all of whom have published fairness-related papers in top-tier venues, contributed their expertise and acted as arbitrators, ensuring the resolution of any disagreements and maintaining the integrity of the analysis.

Finally, to ensure the integrity and reliability of the survey's findings, all authors independently double-checked the content. This review process aimed to identify any potential errors, inconsistencies, or omissions. 

Additionally, as mentioned before, to further enhance the quality of the survey, we shared the draft with the authors of the collected papers. This collaborative approach allowed us to gather valuable input, feedback, and validation from experts in the field. By incorporating their insights, we ensured that the survey accurately represented the original papers' findings and perspectives, strengthening the overall credibility and robustness of our analysis.


\section{Fairness Testing Workflow}\label{workflowsection}
We first introduce existing techniques that support the key activities involved in fairness testing, i.e., test input generation and test oracle identification. 

\begin{table*}[!tp]
\scriptsize
\centering
\caption{Test generation techniques and the fairness definitions adopted.}
\label{testg_summary}
\begin{tabular}{llll}
\hline
Category & Authors [Ref], Year & Venue & Fairness definitions\\
\hline
\multirow{2}*{\tabincell{l}{Random\\generation}} & Galhotra et al. \cite{sigsoftGalhotraBM17,sigsoftAngellJBM18}, 2017 & ESEC/FSE & causal fairness, statistical parity \\
& Angell et al. \cite{sigsoftAngellJBM18}, 2018 & ESEC/FSE & causal fairness, statistical parity \\
\hline
\multirow{18}*{\tabincell{l}{Search-based\\generation}}& Udeshi et al. \cite{kbseUdeshiAC18}, 2018 & ASE & causal fairness\\
& Sano et al. \cite{sekeSanoKT22}, 2022, & SEKE & causal fairness \\
& Aggarwal et al. \cite{sigsoftAggarwalLNDS19}, 2019 & ESEC/FSE & causal fairness\\
& Fan et al. \cite{icseMingFan}, 2022 &  ICSE & causal fairness \\
& Ma et al. \cite{corrabs220908321}, 2022 & arXiv & causal fairness \\
& Zhang et al. \cite{icseZhangW0D0WDD20}, 2020 & ICSE & causal fairness \\
& Zhang et al. \cite{9506918} 2021 & IEEE TSE & causal fairness \\
& Zhang et al. \cite{isstaZhangZZ21}, 2021 & ISSTA & causal fairness\\
& Zheng et al. \cite{icse22fairNEURONFAIR}, 2022 & ICSE & causal fairness\\
& Monjezi et al. \cite{icse23Monjezi}, 2023 & ICSE & causal fairness\\
& Xie and Wu \cite{trustcomXieW20}, 2020 & TrustCom & causal fairness \\
& Perera et al. \cite{esefairnessserart}, 2022 & EMSE & causal fairness \\
& Tao et al. \cite{guanhongfse22}, 2022 & ESEC/FSE & causal fairness\\
& Xiao et al. \cite{isstaxiao23}, 2023 & ISSTA & causal fairness \\
& Patel et al. \cite{icstPatelCLKK22}, 2022 & ICST & causal fairness \\
& Haffar et al. \cite{mdaiHaffarSDJ22}, 2022 & MDAI & counterfactual fairness\\
& Zhang et al. \cite{tosemmengdi}, 2023 & ACM TOSEM & statistical parity \\
& Cabrera et al. \cite{cabrera2019fairvis}, 2019 &  VAST & group fairness \\
\hline
\multirow{4}*{\tabincell{l}{Verification-based\\generation}} & Sharma and Wehrheim \cite{ptsSharmaW20}, 2020 & ICTSS & causal fairness\\
& Sharma et al. \cite{sharma2021mlcheck}, 2021 & ICMLA & causal fairness\\
& Kitamura et al. \cite{ssKitamuraZT22}, 2022 & SSBSE & causal fairness \\
& Zhao et al. \cite{ssbseZhaoTK22}, 2022 & SSBSE & causal fairness \\
\hline
\multirow{18}*{\tabincell{l}{Template-based\\generation}} & D{\'{\i}}az et al. \cite{chiDiazJLPG18}, 2018 & CHI & causal fairness\\
&Zhang et al. \cite{tosemmengdi}, 2023& ACM TOSEM & statistical parity\\
&Liu et al. \cite{colingLiuDFLLT20}, 2020 & COLING & causal fairness, statistical parity\\
& Kiritchenko and Mohammad \cite{starsemKiritchenkoM18}, 2018 & NAACL workshop & causal fairness\\
& Mehrabi et al. \cite{htMehrabiGMPG20}, 2020 &  HT & parity in performance across groups\\
& Wang et al. \cite{aclWangRC22}, 2022 & ACL & parity in performance across groups \\
& Sharma et al. \cite{sharma2021evaluating}, 2020 & NeurIPS workshop & causal fairness \\
& Sheng et al. \cite{emnlpShengCNP19}, 2019 & IJCNLP & causal fairness \\
& Huang et al. \cite{emnlpHuangZJSWRMYK20}, 2020 & EMNLP & causal fairness \\
& Vig et al. \cite{nipsVigGBQNSS20}, 2020 & NeurIPS & counterfactual fairness\\
& Dhamala et al.\cite{fatDhamalaSKKPCG21}, 2021 & FAccT & causal fairness\\
& Smith et al. \cite{emnlpSmithHKPW22}, 2022 & EMNLP & statistical parity\\
& Ribeiro et al. \cite{aclRibeiroWGS20}, 2020 & ACL & causal fairness\\
& Wan et al. \cite{biasaskerfse23}, 2023 & ESEC/FSE & absolute fairness, relative fairness\\
& Ma et al. \cite{ijcai0004WL20}, 2020 & IJCAI & counterfactual fairness \\
& Asyrofi et al. \cite{9653830}, 2021 & IEEE TSE & counterfactual fairness \\
& Yang et al. \cite{sigsoftYangA021}, 2021 & ESEC/FSE & distributional fairness \\
& Sun et al. \cite{icseSunZHPZ20}, 2020 & ICSE & counterfactual fairness\\
& Sun et al. \cite{icse0004Z0HP022}, 2022 & ICSE & counterfactual fairness\\
\hline
Grammar-based generation & Ezekiel et al. \cite{9678017}, 2022 & IEEE TSE & counterfactual fairness \\
\hline
\multirow{7}*{GAN-based generation} & Denton et al. \cite{denton2019image}, 2019 & CVPR workshop & causal fairness \\
& Joo and K{\"{a}}rkk{\"{a}}inen \cite{corrabs200510430}, 2020 & arXiv & causal fairness \\
& Zhang et al. \cite{corabs211108856}, 2021 & arXiv & causal fairness \\
& Muthukumar \cite{Muthukumar19}, 2019 & CVPR workshop & causal fairness \\
& Dash et al. \cite{wacvDashB022} & WACV & counterfactual fairness \\
& Balakrishnan et al. \cite{eccvBalakrishnanXXP20}, 2020 & ECCV & parity in robustness across groups\\
& Pu et al. \cite{icsemetPuKLCL22}, 2022 & ICSE workshop & parity in robustness across groups\\
\hline
\makecell[l]{Signal transformation\\-based generation} & Rajan et al. \cite{faseRajanUC22}, 2022 & FASE & parity in robustness across groups \\ 
\hline
\end{tabular}
\end{table*}

\subsection{Test Input Generation}\label{input_generation}
In the area of fairness testing, test input generation aims to automatically produce instances that can induce discrimination and reveal fairness bugs in software systems. We organize relevant research based on the techniques adopted, including random test input generation, search-based test input generation, verification-based test input generation, and domain-specific test input generation. Table \ref{testg_summary} summarizes these techniques and indicates the fairness definitions that they adopt. 

\subsubsection{Random Test Input Generation}
Random testing evaluates a software system with inputs randomly chosen from its input space, and it can be used to infer the quantitative estimate of a system's operational reliability \cite{tseDuranN84}.

Galhotra et al. \cite{sigsoftGalhotraBM17} and Angell et al. \cite{sigsoftAngellJBM18} introduced Themis, a random test input generation approach for fairness. Themis randomly assigns values to non-sensitive attributes and varies the values of sensitive attributes. By observing the behavior of the system under test with these inputs, Themis quantifies the occurrence of discriminatory instances in the input space. Discrimination is measured using two frequency values. The first value represents the proportion of generated inputs where altering the sensitive attributes leads to a change in the output (\emph{causal fairness}). The second value captures the disparity in receiving favorable outcomes among different demographic groups within the generated inputs (\emph{statistical parity}).

\subsubsection{Search-based Test Input Generation} 
Despite the effectiveness of Themis, random generation can lead to a low success rate of discriminatory input generation \cite{icseMingFan}, so the following fairness testing work generates test inputs using search-based techniques.
Search-based test generation uses meta-heuristic search techniques to guide the generation process and make this process more efficient and effective \cite{infsofHarmanJ01,LakhotiaHM07,icstHarmanJZ15}. It has been employed in an increasing number of fairness testing techniques to explore the input space of the software under test.

\noindent \textbf{Two-phase search-based techniques.} 
In the realm of fairness testing, most search-based test input generation techniques employ a two-phase approach for generating individual discriminatory instances. These techniques operate based on the \emph{causal fairness} definition, which implies that altering sensitive attributes should not impact the outcomes. 

The two phases involved are the global search and the local search:
\begin{itemize}
\item During the \emph{global search} phase, the technique conducts an exploration of the input space to identify an initial set of individual discriminatory instances. These instances consist of pairs where altering the sensitive attributes results in divergent outcomes.
\item During the \emph{local search} phase, the technique focuses on searching for additional individual discriminatory instances in the neighborhood of those found during the global search. This phase is based on the hypothesis that if a discriminatory input exists in the input space, there exist more discriminatory inputs located closer to it \cite{kbseUdeshiAC18}. The hypothesis draws inspiration from the robustness property of ML models, where similar inputs should yield similar outputs. Hence, the discriminatory inputs and their neighbourhood are likely to be similarly discriminatory, especially for robust models \cite{kbseUdeshiAC18}. 
\end{itemize}
In the following, we introduce how existing search-based fairness testing techniques materialize the two phases.

Udeshi et al. \cite{kbseUdeshiAC18} introduced Aequitas, the first fairness testing approach based on a two-phase search framework. In the global search phase, Aequitas randomly explores the input space to uncover discriminatory instances. In the local search phase, Aequitas perturbs the non-sensitive attribute of the discriminatory instances found in the global phase to explore their neighboring inputs and identify more discriminatory instances. Furthermore, Aequitas employs the generated cases to estimate the proportion of inputs that violate causal fairness, offering statistical evidence for fairness bugs.
Sano et al. \cite{sekeSanoKT22} proposed KOSEI, which modifies the local search of Aequitas \cite{kbseUdeshiAC18}. KOSEI individually applies perturbations to all non-protected attributes, in contrast to Aequitas which probabilistically selects one attribute at a time. Additionally, KOSEI enables users to set an iteration limit for the perturbation process.

Aggarwal et al. \cite{sigsoftAggarwalLNDS19} proposed SG, which combines symbolic generation and local explainability for search-based discriminatory instance generation. In the global search phase, SG utilizes a local model explainer to approximate the decision-making process of ML software by constructing a decision tree. Symbolic execution is then employed to cover various paths in the decision tree, aiming to search for discriminatory inputs. In the local search phase, SG perturbs the non-sensitive attribute of these discovered inputs to explore their neighborhood within the input space, thereby generating additional discriminatory inputs.

Fan et al. \cite{icseMingFan} introduced ExpGA, an explanation-guided approach for generating discriminatory instances. Initially, ExpGA employs interpretable methods to identify seed instances that are more likely to produce discriminatory instances when their feature values are slightly modified compared to other instances. Subsequently, using these seed instances as inputs, ExpGA leverages a genetic algorithm for local search, enabling efficient generation of a substantial number of discriminatory instances.

Ma et al. \cite{corrabs220908321} proposed I\&D to enhance the initial seed selection for the global search of the two-phase search-based generation approach. It achieves this by generating a chiral model, which alters the protected attributes of the training data. This chiral model helps identify initial discriminatory instances by detecting differences in predictions compared to the original model. I\&D employs the SHAP value \cite{nipsLundbergL17}, a game-theory-based approach, to explain the variation in prediction behavior between the chiral and original models for each initial discriminatory instance. Moreover, it clusters the initial discriminatory instances based on their SHAP values and selects diverse instances from each cluster in a round-robin fashion for future use in the global search.

In addition to these general techniques, there have been several \emph{two-phase search-based generation techniques specifically proposed for Deep Neural Networks (DNNs)}, including ADF \cite{icseZhangW0D0WDD20,9506918}, EIDIG \cite{isstaZhangZZ21}, NeuronFair \cite{icse22fairNEURONFAIR}, and DICE \cite{icse23Monjezi}.

Zhang et al. \cite{icseZhangW0D0WDD20,9506918} proposed ADF, a gradient-guided generation approach for DNNs. In the global search phase, ADF locates the discriminatory instances near the decision boundary by iteratively perturbing a seed input towards the decision boundary with the guidance of gradient. In the local search phase, ADF further uses gradients as the guidance to search the neighborhood of the found individual discriminatory instances to discover more discriminatory instances.

Zhang et al. introduced EIDIG \cite{isstaZhangZZ21}, which inherits and improves ADF by integrating a momentum term into the iterative search for identifying discriminatory instances. The momentum term enables the memorization of the previous trend and helps to escape from local optima, which ensures a higher success rate of finding discriminatory instances. In addition, EIDIG reduces the frequency of gradient calculation by exploiting the prior knowledge of gradients to accelerate the local search phase. 

Zheng et al. \cite{icse22fairNEURONFAIR} proposed NeuronFair, which uses the identified biased neurons to guide the generation of discriminatory instances for DNNs. In the global search phase, NeuronFair identifies the biased neurons that cause discrimination via neuron analysis and searches for discriminatory instances with the optimization object of increasing the ActDiff (activation difference) values of the biased neurons. In the local search phase, NeuronFair uses the generated discriminatory instances as seeds and perturbs non-sensitive attributes of them to generate more discriminatory instances.

Monjezi et al. \cite{icse23Monjezi} introduced DICE, an information-theoretic approach for fairness testing of DNNs. In the global search phase, DICE uses gradient-guided clustering to explore the input space and identify instances with the maximum quantitative individual discrimination. This discrimination is quantified as the dependence on sensitive attributes using information-theoretic principles. In the local phase, promising instances from the global phase are used to generate additional discrimination instances by exploring their neighborhood.

\noindent \textbf{Other search-based techniques.} Besides the two-phase approach, there are also other search-based generation techniques that have been proposed in the fairness testing literature.

Xie and Wu \cite{trustcomXieW20} used reinforcement learning to develop a black-box search strategy for generating instances that violate \emph{causal fairness}. Their approach frames the task of generating discriminatory instances as a reinforcement learning problem, with the ML model under test treated as part of the environment. The reinforcement learning agent interacts with the environment by taking actions to search for discriminatory inputs, while receiving feedback in the form of rewards and observing the resulting state. Through iterative interactions, the agent learns an optimal policy for efficiently generating discriminatory inputs.

Perera et al. \cite{esefairnessserart} presented SBFT, a search-based fairness testing approach for \emph{regression-based ML models}. SBFT measures unfairness degree as the maximum difference in regression results between pairs of instances that only differ in terms of their sensitive attributes (i.e., \emph{causal fairness}).SBFT begins with an initial set of randomly selected test inputs from the input space. Genetic algorithms are then applied to these inputs, aiming to generate the next generation of test inputs that exhibit the desired fairness degree.

Tao et al. \cite{guanhongfse22} introduced RULER, a fairness testing technique that departs from the strict definition of \emph{causal fairness}. Unlike existing approaches that require coupling samples differing only in sensitive attributes, RULER relaxes this constraint. It allows simultaneous perturbations on both sensitive and non-sensitive attributes to search for more discriminatory instances, because there may not exist discriminatory instances that strictly satisfy the causal fairness. RULER imposes perturbation constraints: sensitive attributes must remain within their valid value ranges, while non-sensitive attributes are bounded by a small range. 

Xiao et al. \cite{isstaxiao23} proposed LIMI, an approach for generating natural discriminatory instances based on \emph{causal fairness}. LIMI uses a generative adversarial network (GAN) to imitate the decision boundary of the model under test in the latent space. By approximating the decision boundary with a surrogate linear boundary,  LIMI can search for instances that closely align with the original data distribution. LIMI performs vector manipulations on latent vectors to move them towards the surrogate boundary. Vector calculations then identify two potential discriminatory candidates in close proximity to the real decision boundary.

Patel et al. \cite{icstPatelCLKK22} applied combinatorial t-way testing \cite{kuhn2013introduction} to fairness testing. Combinatorial t-way testing is a coverage-based data sampling method that can generate diverse datasets by applying logical constraints to specify the sampling space. The approach creates an input parameter model from the training data set and uses the model to generate a t-way test set. For each test, it mutates protected attributes to search for discriminatory instances. 

Haffar et al. \cite{mdaiHaffarSDJ22} introduced two distinct generation approaches for tabular data and images. First, they employed guided adversarial generation on tabular data to search for counterfactuals. Second, they employed GANs to generate counterfactuals for image data.
The generation objective is to modify the predicted label with minimal adjustments to the input. If one of the modified attributes happens to be a protected attribute, it suggests that the model is exhibiting unfairness against that particular attribute.

Zhang et al. \cite{tosemmengdi} introduced TestSGD, a group fairness testing approach for DNNs that specifically focuses on the \emph{statistical parity} of subgroups defined by conjunctions of protected attributes. TestSGD first generates rule sets to capture frequent subgroups. Each rule splits the input space into two groups. To accurately measure group discrimination, TestSGD introduces small, uniform perturbations to the original training samples to search for more samples. Then, it calculates the demographic parity score based on the generated samples.

Cabrera et al. \cite{cabrera2019fairvis} also focused on fairness testing of subgroups and  proposed a testing tool named FAIRVIS. In real-world applications, the number of subgroups to analyze can be overwhelming. To address this challenge, FAIRVIS offers methods to search this large space and find potential issues more efficiently. It introduces a subgroup generation technique that recommends intersectional groups where a model may be underperforming. This technique involves clustering the training dataset to identify statistically similar subgroups and using an entropy-based approach to determine important features and calculate group fairness metrics for the clusters. Finally, FAIRVIS presents generated subgroups sorted by their group fairness metrics.

\subsubsection{Verification-based Test Input Generation}
Since our survey focuses on fairness testing, we do not discuss work focusing solely on formal verification of fairness properties (see e.g., \cite{AlbarghouthiDDN17,pacmplBastani0S19,uaiJohnVS20,pacmplUrbanCWZ20,fmSunSDZ21,aaaiGhoshBM21,aaai22veri}).
Nevertheless, there is a research stream focusing on generating discriminatory test inputs through fairness verification using Satisfiability Modulo Theories (SMT) solving. These techniques are called verification-based test input generation \cite{sharma2021mlcheck}.

Sharma et al. \cite{ptsSharmaW20,sharma2021mlcheck} introduced fairCheck and MLCheck, two verification-based techniques for generating test inputs to assess fairness. These approaches approximate the black-box model under test with a white-box model, leveraging its predictions. The fairness property and the white-box model are translated into logical formulas using SMT solvers. Test cases are subsequently automatically generated in an attempt to violate the specified fairness property, employing the Z3 SMT solver \cite{tacasMouraB08} to check for satisfiability.

Kitamura et al. \cite{ssKitamuraZT22} developed VBT-CT, a technique that integrates combinatorial t-way testing into verification-based fairness testing. 
As described before, combinatorial t-way testing can generate diverse datasets by applying logical constraints to specify the sampling space. 
By incorporating combinatorial t-way testing into the generation process of verification-based testing, VBT-CT enhances the detection capability of discriminatory data.

Zhao et al. \cite{ssbseZhaoTK22} introduced VBT-X, an approach that integrates hash-based sampling \cite{chakraborty2015parallel} into the test generation phase of verification-based fairness testing. Hash-based sampling techniques are capable of producing diverse solutions for a given logical formula, offering the advantage of generating varied solutions with reasonable computational overhead. By leveraging the diverse sampling capability of hash-based sampling, VBT-X enhances the effectiveness of test generation in verification-based fairness testing.

\subsubsection{Domain-specific Test Input Generation}\label{domaingeneration}
Recently, an increasing number of approaches have been proposed for test input generation in specific application domains. These approaches aim to generate natural inputs that belong to the data distribution of a practical application scenario.
This section introduces such domain-specific test input generation in typical domains: natural language processing, computer vision, and speech recognition.

\noindent \textbf{Natural language processing.} Test input generation for Natural Language Processing (NLP) systems is primarily based on that a fair NLP system should produce similar results for pairs of similar texts. 

Researchers can collect text data from the wild and then mutate words related to sensitive attributes to generate test inputs. For example, D{\'{\i}}az et al. \cite{chiDiazJLPG18} conducted a study where they collected sentences containing the word ``old'' from a blogger and replaced it with ``young'' to detect age-related fairness issues in sentiment analysis systems. They also applied mutations to common adjectives instead of age-related words. By using word embeddings, they obtained analogs for ``older'' and ``younger'' versions of these adjectives. These variants were used as test inputs to identify fairness issues related to age. Zhang et al. \cite{tosemmengdi} generated new samples by substituting sensitive terms in the collected texts with alternative terms within the same sensitive attribute category. For instance, for the gender category, the sensitive terms can be bisexual, female, gay, etc. Then they calculate the statistical parity based on the original samples and the generated ones.

Similarly, Liu et al. \cite{colingLiuDFLLT20} conducted fairness testing on generative and retrieval dialogue models. They created gender and race word lists, including male-female word pairs and African American English-Standard English word pairs. From a large dialogue corpus, they selected contexts containing words from these lists and created parallel contexts by replacing the words with their counterparts. The original and generated texts were then compared based on the response of dialogue models using four measurements: diversity, politeness, sentiment, and attribute words. The attribute word analysis involved comparing the probability of attribute words (e.g., career and family-related) appearing in the responses across different groups' contexts. This is a mixed adoption of causal fairness and statistical parity.

\textbf{\emph{Generation based on handcrafted templates:}} A large proportion of fairness testing studies for NLP systems involve the use of handcrafted templates to generate test inputs. These templates consist of short sentences with placeholders (e.g., ``\emph{$<$person$>$ goes to the school in our neighborhood}'') that can be filled with different words to test for violations of causal fairness. Due to the simplicity of the handcrafted templates, most of such test generation studies adopt \emph{causal fairness} that detects whether NLP systems produced similar outcomes for texts that differ only in sensitive attributes.

Kiritchenko and Mohammad \cite{starsemKiritchenkoM18} created 11 templates focused on gender and race, using predefined values for the placeholder \emph{$<$person$>$} representing different names and noun phrases referring to females/males or African/European American. Then they compared emotion and sentiment scores that the systems predict on pairs of sentences differ only in one word corresponding to race or gender.\footnote{Sex and gender are different concepts that are often used interchangeably. We keep the original usage of the two words in each paper to reserve fidelity.}

Mehrabi et al. \cite{htMehrabiGMPG20} created nine templates to compare the performance of name entity recognition systems with respect to recognizing male and female names. Wang et al. \cite{aclWangRC22} focused on machine translation systems and developed 30 templates to test their ability to determine the correct gender of a name. 

Sharma et al. \cite{sharma2021evaluating} designed three templates specifically for gender-related fairness issues in natural language inference systems, with gender-specific hypotheses using the placeholder \emph{$<$gender$>$}.

Similarly, other researchers have designed templates for detecting fairness bugs in natural language generation and machine translation systems. Sheng et al. \cite{emnlpShengCNP19}, Huang et al. \cite{emnlpHuangZJSWRMYK20}, Vig et al. \cite{nipsVigGBQNSS20}, and Dhamala et al. \cite{fatDhamalaSKKPCG21} used templates with placeholders for sensitive attributes such gender and race. 

Smith et al. \cite{emnlpSmithHKPW22} developed a set of manually created templates covering 13 different demographic attributes. These templates were used to identify bias in language models by examining \emph{statistical disparity}, which measures the differences at the group level in the output or assigned probabilities of the model, which arise due to the presence of different identity or demographic information within the input text.

Another approach, CheckList \cite{aclRibeiroWGS20}, utilizes predefined templates to evaluate NLP systems on various capabilities, including fairness.

Wan et al. \cite{biasaskerfse23} introduced BiasAsker, a test input generation approach aimed at measuring absolute bias and relative bias of conversational systems towards various demographic groups. Absolute bias refers to direct expressions of bias, such as statements like ``Group A is smarter than Group B,'' while relative bias involves generating different responses to questions about different groups. To obtain social groups and biased properties, they constructed a comprehensive social bias dataset, which includes a total of 841 groups and 8,110 biased properties. To detect both types of bias, they designed templates and rules for generating Yes-No Questions, Choice-Questions, and Wh-Questions. These generated inputs serve as a means to evaluate and identify bias in conversational systems.

\textbf{\emph{Automated generation:}} While handcrafted templates have been effective in detecting fairness issues in NLP systems, researchers argue that the generated test inputs relying on them may be simplistic and limited \cite{abs200500357}. Furthermore, the generated simplistic test inputs may overlook complex scenarios where multiple words related to sensitive attributes are present, which is often the case in practical applications. To address this limitation, automated test generation techniques should carefully consider each word that can rely on sensitive attributes and generate natural counterfactual inputs that align with the \emph{counterfactual fairness} definition.

To address the problem, Ma et al. \cite{ijcai0004WL20} introduced the automated framework MT-NLP. It employs advanced NLP techniques to identify human-related tokens in the input text and utilizes word analogy techniques to mutate these tokens, generating discriminatory test inputs. Language fluency metrics are then used to filter out unrealistic inputs.

Asyrofi et al. \cite{9653830} proposed BiasFinder, an automated approach for creating diverse and complex test inputs to uncover fairness bugs in NLP systems. By leveraging NLP techniques like coreference resolution and named entity recognition, BiasFinder identifies words associated with demographic characteristics and replaces them with placeholders to form templates. Concrete values are then filled into these templates, generating a large number of text mutants for testing metamorphic relationships. 

Yang et al. \cite{sigsoftYangA021} proposed BiasRV, an approach for testing bias in deployed sentiment analysis systems. By utilizing BiasFinder, they generate a template for a given text and create mutants based on that template to assess if the system exhibits biased predictions. They define \emph{distributional fairness}, which examines whether mutants from different demographic groups are treated similarly. Specifically, they expect the distribution of predicted sentiments for these groups of mutants to be closely aligned if the system is fair. Unlike traditional statistical parity, which measures overall fairness, distributional fairness evaluates whether the system makes biased predictions for specific inputs.

Ezekiel et al. \cite{9678017} developed ASTRAEA, a grammar-based fairness testing approach for generating discriminatory inputs in NLP systems. ASTRAEA incorporates input grammars covering various NLP tasks and biases. By randomly exploring the input grammars and mutating sensitive attribute-related words using alternative tokens, ASTRAEA generates initial test inputs and checks their satisfaction of metamorphic relations.

For fairness testing of machine translation systems, Sun et al. \cite{icseSunZHPZ20,icse0004Z0HP022} proposed TransRepair and CAT. TransRepair conducts sentence mutations by replacing words with context-similar alternatives, while CAT identifies and replaces words using isotopic replacement. The resulting mutants, along with the original input sentence, serve as test inputs for evaluating the fairness of machine translation.

\noindent \textbf{Computer vision.}
To detect fairness issues in Computer Vision (CV) systems, researchers often examine how the system's output changes when the sensitive attribute of a person in the input image is altered while keeping other factors constant (i.e., generating \emph{counterfactual} images). This idea forms the basis of test input generation for CV systems.

GANs \cite{GoodfellowPMXWOCB14} are commonly used for image transformations in ML testing \cite{tseZhangHML22}. However, conventional GANs face challenges in generating precise changes required for fairness testing. For example, changing hair color without affecting other facial features or hair style can be difficult.
To address this, recent efforts have adapted and improved conventional GANs for generating test inputs in CV systems. Denton et al. \cite{denton2019image} developed a face-generative model that maps latent codes to images and inferred directions in latent code space to manipulate specific sensitive attributes. Test inputs were generated by traversing these inferred directions. Joo and K{\"{a}}rkk{\"{a}}inen \cite{corrabs200510430} employed the FaderNetwork architecture \cite{nipsLampleZUBDR17}, where specific known attributes of an image are input separately to the generator. Zhang et al. \cite{corabs211108856} utilized CycleGAN \cite{iccvZhuPIE17}, which limits changes to non-sensitive attributes, to generate discriminatory inputs. 

Muthukumar \cite{Muthukumar19} criticized GAN-based approaches for their inability to effectively modify a single attribute while keeping other non-related attributes unchanged. This limitation makes it challenging to identify the exact cause of unfair outcomes. For instance, in gender classification where discrimination occurs between different skin types, it is possible that factors like hairstyle, facial structure, cosmetics, or clothing contribute to the disparity rather than skin type alone. To address this issue, Muthukumar proposed a solution where face images were represented in the YCrCb color space \cite{pamiHsuAJ02}, and techniques such as luminance mode-shift and optimal transport \cite{FerradansPPA14} were utilized to alter the skin type of a face.

However, these approaches may fail to account for causal relationships between attributes when generating discriminatory images. Although they claim to produce counterfactual images, they often rely on a causal fairness definition rather than strict counterfactual fairness. To generate test inputs that reflect real-world scenarios, it is crucial to consider the downstream effects resulting from changes in sensitive attributes. For instance, in a chest MRI classification system, a patient's age can influence the relative size of their organs \cite{wacvDashB022}. Therefore, altering the age without considering the causal relationship between age and organ size would lack realism. In response to this limitation, Dash et al. \cite{wacvDashB022} introduced ImageCFGen, a fairness testing method that incorporates knowledge from a causal graph and utilizes an inference mechanism within a GAN-like framework. This approach enables the generation of discriminatory images that adhere to the definition of counterfactual fairness.

Unlike the aforementioned studies that employ individual fairness definitions, Balakrishnan et al. \cite{eccvBalakrishnanXXP20}  introduced a new concept of group fairness, investigating the parity in robustness against transformations across various demographic groups. Specifically, they proposed an image generation method that produces synthesized samples considering multiple sensitive attributes, and compares error rates of gender classifiers across various subgroups before and after the synthesis. For generation, the approach modifies multiple attributes simultaneously to create grid-like sets of matched images called transects. This is achieved by navigating the latent space of the generator in directions specific to each attribute.

Similarly, Pu et al. \cite{icsemetPuKLCL22} investigated the parity in robustness of deepfake detectors by introducing makeup, as a feature perturbation across various demographic groups. Specifically, they employed common makeup alterations such as eyeshadows, eyeliners, lipstick, and blushes to perturb the face images before feeding these images to the deepfake detection models. Subsequently, they compared the accuracy differences of the detectors in detecting deepfakes between original and synthetic images for both male and female images, thus quantifying the extent of gender disparity in the two accuracy differences. In addition, they discovered that salient regions near the lips have the greatest impact on the fairness of the tested models.

\noindent{\textbf{Speech recognition.} The fairness testing literature has dedicated less attention to speech recognition compared to NLP and CV. Rajan et al. \cite{faseRajanUC22} introduced AequeVox, a fairness testing approach tailored for speech recognition systems. Similar to Balakrishnan et al. \cite{eccvBalakrishnanXXP20} and Pu et al. \cite{icsemetPuKLCL22}, AequeVox measures group fairness as the parity in robustness against perturbations across demographic groups. It generates test inputs by applying eight common real-life metamorphic perturbations to speech signals, such as noise, drop, and low/high pass filters. Then it calculates the error rate increase in speech recognition for various demographic groups after applying these perturbations. If the difference in error rate increase surpasses a predetermined threshold for different groups, the recognition system is considered to have fairness bugs.

\subsubsection{Test Input for Data Testing}
ML program testing and model testing typically rely on data instances as test inputs, whose generation process has been detailed in the previous sections. For data testing, test inputs are typically programs capable of detecting data bias, as described in Section~\ref{ftworkf}. Typically, researchers manually implement these programs based on testing objectives, which include identifying feature bias, label bias, and selection bias. 

To detect feature bias,  Peng et al. \cite{peng2021xfair} implemented logistic regression and decision tree algorithms as models to infer relationships between sensitive attributes and non-sensitive features. This approach facilitated the identification of non-sensitive features that are correlated with sensitive attributes and contribute to fairness issues. Li et al. \cite{icse22fair2} implemented linear regression models to analyze the association between non-sensitive features and sensitive attributes to identify feature bias. Zhang et al. \cite{sbrimsZhangWW16,ijcaiZhangWW17,tkdeZhangWW19} implemented a causal graph to identify features causally leading to unfair outcomes. Li and Xu \cite{iccvLiX21} optimized a hyperplane within the latent space of a generative model, offering a solution to detect unknown biased attributes and enhance the understanding of feature bias. Black et al. \cite{fatBlackYF20} developed optimal transport projection programs \cite{villani2009optimal} to map data instances from the privileged/unprivileged group to their counterparts in the unprivileged/privileged group, conducting analyses on these flipsets to detect features contributing to unfairness. 

To detect label bias, Chakraborty et al. \cite{sigsoftChakrabortyMM21,sigsoftChakrabortyM0M20,fairsslware} implemented situation testing programs to  uncover biased data instances. Additionally, Chen and Joo \cite{iccvChenJ21} tailored testing programs based on facial action units to detect label bias in widely-used datasets for facial expression recognition.

To detect selection bias,  researchers \cite{zhenpengmaat22,sigsoftChakrabortyMM21,fairsslware} implemented programs to analyze the statistical association between sensitive attributes and outcome labels in the training data. Simultaneously, some researchers \cite{datasetbiasicpr} scrutinized dataset representation to ascertain potential underrepresentation issues of demographic groups.

\subsection{Test Oracle Identification}\label{oracle}
Given an input for a software system, the challenge of distinguishing the corresponding desired behavior from the potentially incorrect behavior is called the ``\emph{test oracle problem}'' \cite{tseBarrHMSY15}. Test oracle identification is one of the key problems in ML testing \cite{tseZhangHML22}. The test oracle of fairness testing determines whether the software is behaving as fairness requirements and enables the judgment of whether a fairness bug exists. Existing work employs two types of test oracles for fairness testing: metamorphic relations and statistical measurements.

\subsubsection{Metamorphic Relations as Test Oracles}\label{Metamorphicrelations}
A metamorphic relation is a relationship between the software input change and the output change that we expect to hold across multiple executions \cite{tseBarrHMSY15}. Suppose a system that implements the function $\sin(x)$, then $\sin(x)=sin(\pi+x)$ is a metamorphic relation. This relation can be used as a test oracle to help detect bugs. If $\sin(x)$ differs from $sin(\pi+x)$, we can conclude that the system under test has a bug without the need for examining the specific values output by the system. 

Metamorphic relation is a type of pseudo oracle commonly adopted to automatically mitigate the oracle problem in ML testing. It has also been widely studied for fairness testing of ML software. Specifically, existing work mainly performs fairness-related metamorphic transformation on the input data or training data of ML software, and expects these transformations to not change or yield expected changes in the prediction. Next, we classify and discuss this work according to whether the metamorphic transformations operate on sensitive attributes.

\noindent \textbf{Metamorphic transformations through mutating sensitive attributes.} 
The mutation of sensitive attributes is a widely used technique for generating metamorphic relations in the context of \emph{individual fairness}. Assessing violations of individual fairness, such as counterfactual fairness and causal fairness, typically requires the comparison of paired data instances that vary in their sensitive attributes.

For ML software designed for \emph{classification} tasks, a widely employed metamorphic relation for fairness testing involves comparing pairs of instances with different sensitive attributes but similar non-sensitive attributes, expecting them to yield the same classification outcome. For instance, a fair loan application system should make identical decisions for two applicants who differ only in their gender.

This metamorphic relation has found extensive application in testing the fairness of software systems and guiding the generation of fairness tests across various domains, including tabular data classification \cite{sigsoftGalhotraBM17,isstaZhangZZ21,9506918,icseZhangW0D0WDD20,trustcomXieW20,sigsoftAggarwalLNDS19,kbseUdeshiAC18,sigsoftAngellJBM18}, text classification \cite{9678017,9653830,starsemKiritchenkoM18,aclRibeiroWGS20,aclWangRC22,sharma2021evaluating,emnlpHuangZJSWRMYK20,ijcai0004WL20}, and image classification \cite{icse22fairNEURONFAIR,corabs211108856,wacvDashB022,corrabs190606439,corrabs200510430,denton2019image}, among others.

In the case of tabular data, researchers typically select the sensitive attributes of interest in the dataset (e.g., gender and race) and modify the values of those attributes (e.g., from ``male'' to ``female'') to assess if the software violates the fairness metamorphic relation for those instances (i.e., adopting the \emph{causal fairness} definition). Tao et al. \cite{guanhongfse22} proposed that altering only the sensitive attributes can lead to unnatural test inputs and therefore relaxed this constraint by allowing small permutations to be applied to non-sensitive attributes simultaneously. Similarly, counterfactual fairness \cite{nipsKusnerLRS17} necessitates modifying the non-sensitive attributes that are causally influenced by the sensitive attributes whenever the sensitive attributes are modified.

For text data, researchers generate inputs by filling sensitive attribute-related placeholders in predefined templates and mutating the sensitive attributes within these templates, or identify the entities related to the sensitive attributes and simultaneously transform all the identified entities to create valid test inputs that reflect real-world scenarios. For example, when dealing with gender, researchers identify and modify person names, gender pronouns, gender nouns, and so on.

For image data, researchers predominantly rely on advanced deep learning techniques such as GANs \cite{GoodfellowPMXWOCB14} to transform images across different sensitive attributes. The methodologies for altering sensitive attributes in different types of data are extensively discussed in Section \ref{input_generation}.

For ML software designed for \emph{regression} tasks, the prediction outcomes are continuous values rather than discrete labels, which poses a challenge in determining metamorphic relations for assessing fairness. Specifically, it is difficult to ascertain whether two predicted continuous outcomes are different enough to indicate fairness issues in the software under test. To address this challenge, Udeshi et al. \cite{kbseUdeshiAC18} introduced the use of a threshold to define metamorphic relations. In this approach, the difference between the outcomes of two similar instances that vary only in their sensitive attribute must be smaller than a manually specified threshold. Similarly, Perera et al. \cite{esefairnessserart} proposed the concept of fairness degree, which quantifies the maximum difference in predicted values across all pairs of instances that are similar except for their sensitive attribute. The fairness degree can be utilized and defined to construct metamorphic relations and guide the generation of test inputs.

For ML software designed for \emph{generation} tasks, it is also challenging to determine the metamorphic relations. In the case of natural language generation systems, it is particularly difficult to evaluate the identity or similarity of generated text. To address this, researchers \cite{emnlpHuangZJSWRMYK20,emnlpShengCNP19,colingLiuDFLLT20,fatDhamalaSKKPCG21} have employed various existing natural language processing techniques to measure text similarity. These techniques include sentiment classification, perplexity and semantic similarity measurement, politeness measurement, diversity measurement, toxicity classification, and regard classification applied to machine-generated text. As a result, metamorphic relations for text generation systems require that pairs of inputs, which are identical except for the sensitive attribute, should yield generated text with consistent sentiment polarity, perplexity, semantics, and regard. This serves as a measure of similarity in the generated text.
In the context of machine translation systems, where translations are generated based on input sentences, Sun et al. \cite{icseSunZHPZ20,icse0004Z0HP022} have proposed generating test oracles based on the metamorphic relationship between translation inputs and outputs. Specifically, they expect translation outputs for the original input sentence and its mutants, considering sensitive characteristics, to exhibit a certain level of consistency modulo the mutated words. To validate this consistency, similarity metrics are employed as test oracles to measure the degree of agreement between translated outputs.

\noindent \textbf{Metamorphic transformations through mutating non-sensitive attributes.} Some researchers \cite{icsemetPuKLCL22,faseRajanUC22} adopt a different approach to generate metamorphic relations for fairness testing by focusing on the mutation of non-sensitive attributes. They employ metamorphic transformations as perturbations applied to samples and subsequently assess the ML software's robustness to these perturbations across various demographic groups. This methodology can be viewed as an implementation of group fairness, as it involves comparing the software's performance across different groups in response to the perturbations.

Balakrishnan et al. \cite{eccvBalakrishnanXXP20}  produced synthesized samples taking multiple sensitive attributes into account, and compared error rates of gender classifiers across various subgroups before and after the synthesis.

Pu et al. \cite{icsemetPuKLCL22} employed makeup as a form of perturbation applied to face images. They then compared the accuracy differences as bias factors between male and female individuals on both the original face images and synthetic images generated by introducing these perturbations.

Rajan et al. \cite{faseRajanUC22} applied eight metamorphic transformations to speech signals, and measured the increase in error rates of speech recognition for different demographic groups after these transformations.

\subsubsection{Statistical Measurements as Test Oracles}\label{statistical_oracle}
Researchers have proposed various statistical fairness measurements that align with different fairness definitions. While these measurements do not serve as direct oracles for fairness testing, they provide a quantitative way to assess the fairness of the software under test.
For instance, in the case of statistical parity, researchers calculate the favorable rate among demographic groups and identify fairness violations by comparing these rates. This comparison involves measuring the difference between the rates, known as Statistical Parity Difference (SPD), or computing the ratio of the rates, known as Disparate Impact (DI) \cite{icseZhangH21}.
If the calculated SPD or DI exceeds a predefined threshold, it indicates the presence of fairness bugs in the software under test.

There is a wide array of statistical fairness measurements available, with the IBM AIF360 toolkit \cite{AIF360link} alone offering over 70 such measurements. Determining the appropriate measure of fairness is a requirements engineering problem involving negotiation among various stakeholders and different interpretations\cite{baresiREnext}. For certain social-critical application scenarios, domain knowledge is required for fairness testing and the \textit{prediction-modeler} (e.g., data scientists and software engineers) needs to work together with the \textit{decision-maker} (e.g., product managers and business strategists) \cite{SarroRE}. How to achieve this is still an open challenge. However, once a fairness measure is determined, the actual statistical measurement used for testing is typically straightforward.
Providing a comprehensive description and comparison of each measurement is beyond the scope of this survey. Verma and Rubin \cite{conficseVermaR18} conducted a survey and categorized several widely adopted statistical fairness measurements. In this survey, we expand upon their categorization based on our collected papers and present representative measurements from each category.

\noindent \textbf{Measurements based on predicted outcomes.} Some measurements are calculated based on the predicted outcomes of the software for privileged and unprivileged groups. For example, the aforementioned Statistical Parity Difference (SPD) \cite{barocas2016big} measures the difference in favorable rates among different demographic groups; Disparate Impact (DI) \cite{kddFeldmanFMSV15} measures the ratio of the favorable rate of the unprivileged group against that of the privileged group.

\noindent \textbf{Measurements based on predicted and actual outcomes.} Some measurements not only consider the predicted outcomes for different demographic groups, but also compare them with the actual outcomes recorded in the collected data. For example, the Equal Opportunity Difference (EOD)~\cite{nipsHardtPNS16} measures the difference in the true-positive rates of privileged and unprivileged groups, where the true-positive rates are calculated by comparing the predicted and actual outcomes. Another widely adopted measurement that lies in this category is Average Odds Difference (AOD) \cite{nipsHardtPNS16}, which refers to the average of the false-positive rate difference and the true-positive rate difference between unprivileged and privileged groups.

\noindent \textbf{Measurements based on predicted probabilities and actual outcomes.} Some measurements take the predicted probability scores and actual outcomes into account. For example, for any given predicted score, the calibration measurement calculates the difference in the probability of having a favorable outcome for privileged and unprivileged groups \cite{bigdataChouldechova17}; the measurement of balance for positive class calculates the difference of average predicted probability scores in the favorable class between privileged and unprivileged groups~\cite{inKleinbergMR17}.

\noindent \textbf{Measurements based on neuron activation.} As DNNs are widely used in software systems to support the decision-making process, researchers have started to leverage the internal behaviors of DNNs to design statistical fairness measurements. Tian et al. \cite{icseTianZOKR20} proposed a new statistical measurement based on neuron activation for DNNs. First, they computed a neuron activation vector for each label class based on the test inputs. Specifically, for a class $c$, each element of its neuron activation vector represents how frequently a corresponding neuron is activated by all members in the test inputs belonging to class $c$. Then they computed the distance between neuron activation vectors of different classes as the fairness measurement. If two classes do not show a similar distance with regards to a third class, they consider that the DNN under test contains fairness bugs.

\noindent \textbf{Measurements based on situation testing.} Some researchers designed statistical measurements to approximate situation testing, which is a legal experimental procedure of seeking pairs of instances that have similar characteristics apart from the sensitive attribute value, but obtain different prediction outputs \cite{kddThanhRT11}. Thanh et al. \cite{kddThanhRT11} leveraged the k-nearest neighbor classification to approximate situation testing. They first divided the dataset into the privileged group and the unprivileged group, based on the sensitive attribute. Then, for each instance $r$ in the dataset, they found the k-nearest neighbors in the two groups and denoted them as sets $K_p$ and $K_u$, respectively. Finally, they calculated the proportions of instances, for which the outcome is the same as $r$ in $K_p$ and $K_u$, and measured the difference between the two proportions. If the difference is larger than a given threshold, the instance $r$ is considered unfairly treated. Zhang et al. \cite{ijcaiZhangWW16} improved the measurement proposed by Thanh et al. \cite{kddThanhRT11}. They designed a new distance function that measures the distance between data instances, to improve the k-nearest neighbor classification. Their function considers only the set of attributes that are identified as the direct causes of the outcome by Causal Bayesian Networks \cite{pearl2009causality}.

\noindent \textbf{Measurements based on optimal transport projections.} Several measurements \cite{fatBlackYF20,TaskesenBKN21,icmlSiMBN21} are proposed based on optimal transport projections \cite{villani2009optimal}, which seek for a transformation map between two probability measures. Black et al. \cite{fatBlackYF20} mapped the set of women in the data to their male correspondents, with the optimal transport projection to minimize the sum of the distances between a woman and the man to which she is mapped (called her counterpart). Then they extracted the positive flipset, which contained the women with favorable outcomes whose counterparts did not. They also extracted the negative flipset, which was the set of women with unfavorable outcomes whose counterparts are favorable. Finally, they calculated the size difference of the positive and the negative flipsets to measure the unfairness of the system under test.

\noindent \textbf{Measurements for ranking systems.} Applying the aforementioned statistical fairness measurements directly to ranking systems, which are extensively utilized in various domains such as hiring and university admissions \cite{aiesKuhlmanGR21,irGeziciLSY21}, poses significant challenges. To address this challenge, some researchers have tackled the ranking problem by transforming it into a classification problem and subsequently applying existing statistical fairness measurements. For example, researchers \cite{singh2018fairness,yang2017measuring} have used statistical parity difference as a fairness measurement, assessing whether individuals from different groups have equal representation among desirable outcomes, such as securing top positions in the ranking.
Pairwise fairness is another common statistical metric employed in the context of ranking systems \cite{kuhlman2019fare,NarasimhanCGW20,BeutelCDQWWHZHC19}. It necessitates that a ranking system ensures an equal likelihood of a clicked item being ranked above another relevant unclicked item across different demographic groups.
Recently, a fairness testing approach for deep recommender systems, namely FairRec \cite{isstaFairRec}, supports the measurement of differences in evaluation metrics between demographic groups through three types of statistical measurements: recommendation performance, alignment of recommended item popularity with user preferences, and diversity in recommendations. 

\noindent \textbf{Measurements considering multiple subgroups.} Some researchers extend their analysis to encompass a spectrum of subgroups, representing combinations of various sensitive attributes. This approach allows for a more nuanced examination of fairness in ML software. The consideration of multiple subgroups enhances the scope of fairness measurements, recognizing the diverse perspectives and experiences within different segments of the population. For instance, Chen et al.~\cite{corrabs230801923} used intersectional fairness as a measurement, calculating the maximum disparity between any two subgroups. Similarly, Zhang et al. \cite{tosemmengdi} proposed a group fairness testing approach that specifically focuses on the SPD within subgroups.

It is difficult to determine the threshold for statistical measurements to detect the fairness bugs. For example, for the aforementioned SPD, it would be too strict to consider the software under test to be fair only when SPD equals to 0. In practice, practitioners can set a threshold for the measurement under consideration \cite{kddDiCiccioV0KA20}. If the measurement result for the software under test is above or below the specified threshold, the software is considered to have a fairness bug. Although the threshold could be empirically specified by engineers, it is challenging to determine the appropriate threshold for each fairness measurement. 

To alleviate this problem, researchers attempt to use statistical testing, based on the measurements to detect fairness bugs. Tramèr et al. \cite{TramerAGHHHJL17} proposed FairTest to analyze the associations between software outcomes and sensitive attributes. The software under test is deemed to have a fairness bug, if the associations are statistically significant.  Taskesen et al. \cite{TaskesenBKN21} and Si et al. \cite{icmlSiMBN21} employed a statistical hypothesis test for the fairness measurements based on optimal transport projection.
DiCiccio et al. \cite{kddDiCiccioV0KA20} presented a non-parametric permutation testing approach for assessing whether a software system is fair in terms of a fairness measurement. The permutation test is used to test the null hypothesis that a system has equitable performance for two demographic groups (e.g., male or female) with respect to the given measurement. Gursoy et al. \cite{corrabs220808279} used the permutation test to detect whether prediction errors of a regression model are distributed in a statistically significant manner across demographic groups to determine whether the model under test is unfair.

In addition, researchers construct the baseline for the fairness measurement, and detect fairness bugs by comparing the measurement value with the baseline. Zhao et al. \cite{emnlpZhaoWYOC17} used the fairness measurements calculated based on training data as their baseline against which to evaluate. Specifically, they used the obtained ML model to annotate unlabeled data instances, and revealed situations when the ML process amplified existing bias by comparing the fairness measurements on training data and those on the annotated dataset.
Wang and Russakovsky \cite{icmlWangR21} showed that the bias amplification measurement proposed by Zhao et al. \cite{emnlpZhaoWYOC17} conflated different types of bias amplification and failed to account for varying base rates of sensitive attributes. Then they proposed a new, decoupled metric for measuring bias amplification, which takes into account the base rate of each sensitive attribute and disentangles the directions of amplification.
Wang et al. \cite{iccvWangZYCO19} presented a fairness testing approach for visual recognition systems that predicted action labels for images containing people. They trained two classifiers to predict gender from a set of ground truth labels and model predictions. The difference in the predictability of the two models indicated whether the ML process introduced fairness bugs.

\subsubsection{Test Oracle for Data Testing} ML model testing relies on metamorphic relations and statistical measurements as test oracles. Moreover, all ML program testing papers that we collect use statistical measurements (in Section \ref{statistical_oracle}) as test oracles. This section describes test oracles for data testing of fairness. Specifically, we summarize the test oracles for detecting feature bias, label bias, and selection bias, respectively.

\emph{Feature bias:} Peng et al. \cite{peng2021xfair} and Li et al. \cite{icse22fair2} assumed that feature bias is present when non-sensitive attributes exhibit correlations with sensitive attributes. In contrast, some researchers \cite{sbrimsZhangWW16,ijcaiZhangWW17,tkdeZhangWW19,iccvLiX21} argued that feature bias manifests if specific features establish causal relationships with unfair outcomes. Black et al. \cite{fatBlackYF20} mapped data instances from the privileged/unprivileged group to their counterparts in the unprivileged/privileged group and determined biased features based on divergent outcomes among these counterparts.

\emph{Label bias:} Chakraborty et al. \cite{sigsoftChakrabortyMM21,sigsoftChakrabortyM0M20,fairsslware} partitioned the training data into privileged and unprivileged groups, and then trained two models separately for each group. Subsequently, they employed the two models to predict outcomes for each training data instance, identifying label bias if the predicted outcomes differ.

\emph{Selection bias:} Chen et al. \cite{zhenpengmaat22} and Mambreyan et al. \cite{datasetbiasicpr} assumed that selection bias is present when a statistical association exists between a sensitive attribute and outcome labels in the training data. In contrast, some researchers \cite{KarkkainenJ21, cvprTorralbaE11, aiesYangGFSY0NHJ22,eccvWangNR20} assumed that different demographic groups should be equally represented in the dataset, implying an equal number of data instances for each group. Some scholars combined both oracles, expecting fair training data to exhibit an equivalent favorable rate between privileged and unprivileged groups (i.e., no statistical association between a sensitive attribute and outcomes) and an equal number of data instances for both groups \cite{sigsoftChakrabortyMM21,fairsslware}.

\section{Fairness Testing Components}\label{testcomponent}
This section introduces the fairness testing literature from the perspective of ``\emph{what to test}.'' Just as traditional software testing can be conducted on different testable parts within a software system \cite{LaghariD18}, fairness testing can also be performed on different parts, including training data, ML programs, ML models, ML frameworks, and non-ML components. Fig. \ref{fig:component} shows the categorization of this section. Existing studies primarily focus on testing training data, ML programs, and ML models.

\begin{figure}[t]
    \centering
       \includegraphics[width=0.7\linewidth]{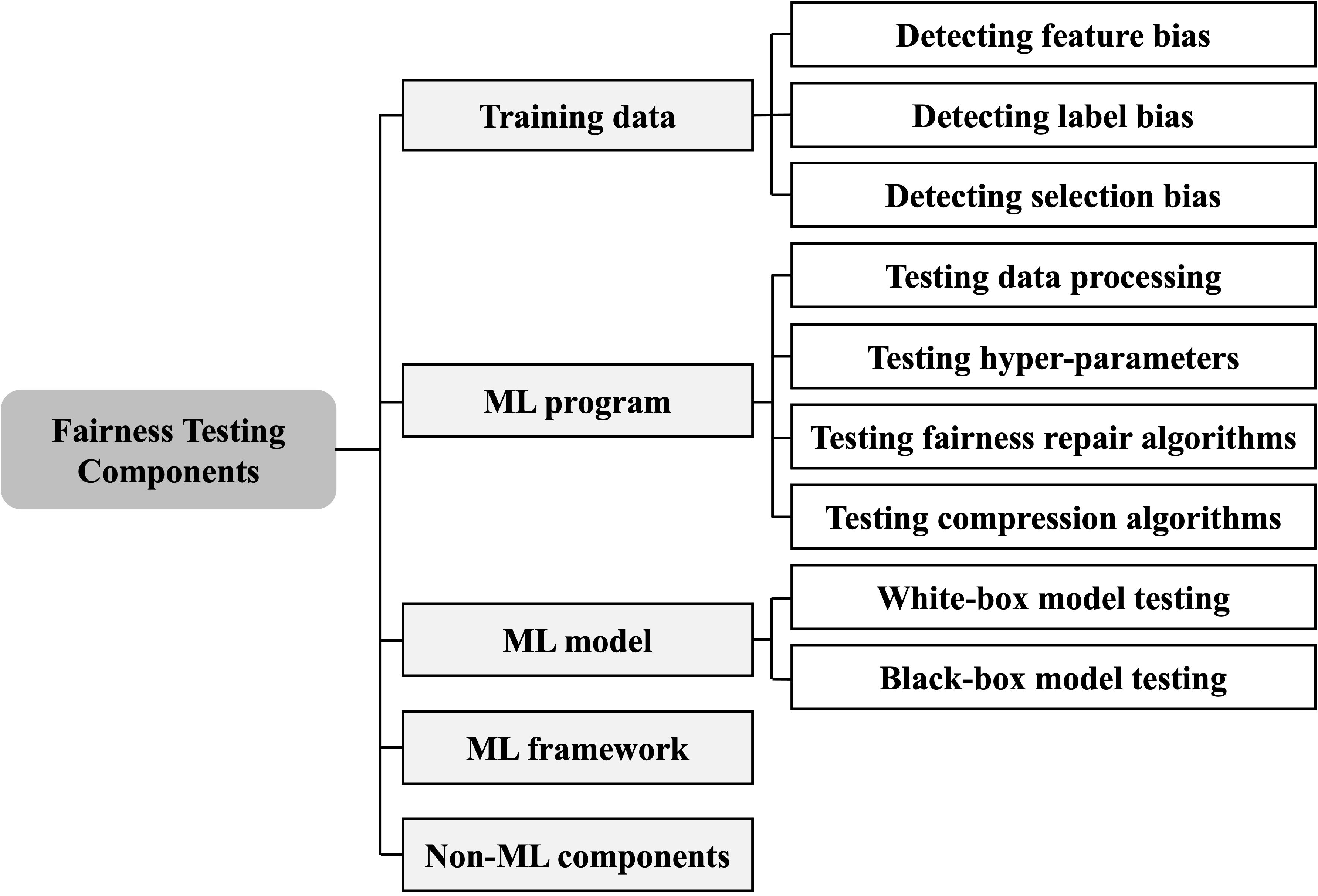}
  \caption{Testing components of fairness testing.}
  \label{fig:component} 
\end{figure}

\subsection{Data Testing}\label{datatesting}
ML models are developed following the data-driven paradigm. This paradigm makes ML models vulnerable to fairness bugs present in data. Specifically, fairness bugs in training data can be learned and propagated throughout the ML model development pipeline, leading to the creation of biased and unfair ML software systems.
To tackle this problem, data testing approaches, which detect bugs in ML training data~\cite{tseZhangHML22}, have been proposed for fairness testing. They detect the bias in data features, data labels, and data distribution.

\subsubsection{Detecting Feature Bias} Feature bias arises when certain features in the training data exhibit a strong correlation with sensitive attributes, causing them to become the underlying source of software unfairness \cite{icse22fair2}. Zhang and Harman \cite{icseZhangH21} investigated the impact of the feature set on the fairness of ML models. Their findings demonstrated that the selection of features has a substantial influence on fairness, thereby highlighting the importance of considering testing data features in fairness-related endeavors.

In order to identify the features responsible for fairness issues, it is natural to suspect that the software exhibits discrimination against a particular demographic group due to its consideration of sensitive attributes during both the training and prediction phases \cite{sigsoftChakrabortyM0M20}. To investigate whether the sensitive attribute serves as the underlying cause of fairness problems, Chakraborty et al. \cite{sigsoftChakrabortyM0M20} conducted an experiment where they removed the sensitive attribute information from the data (i.e., fairness through unawareness). Surprisingly, they found that the resulting machine learning software demonstrated a similar degree of unfairness as observed previously.

A similar discovery was made in a real-world case: back in 2016, it was revealed that Amazon's same-day delivery service exhibited discriminatory behavior towards neighborhoods with a disproportionately high population of Black residents \cite{amazoncase}. Although the ML model behind the service did not explicitly incorporate race information, the presence of correlated attributes in the training data allowed for the possibility of bias. Specifically, it was found that the ``Zipcode'' information utilized during model training exhibited a strong correlation with race, causing the ML model to indirectly infer race information from it.

To identify non-sensitive features that could potentially contribute to fairness issues, Peng et al. \cite{peng2021xfair} employed logistic regression and decision tree algorithms as models to infer the relationships between sensitive attributes and non-sensitive features. Similarly, Li et al. \cite{icse22fair2} utilized linear regression to analyze the association between each feature and sensitive attributes, thereby identifying features that may introduce bias.

In contrast, Zhang et al. \cite{sbrimsZhangWW16,ijcaiZhangWW17,tkdeZhangWW19} employed discrimination detection based on causal modeling to detect both direct and indirect discrimination within datasets. They constructed a causal graph to capture the causal relationships between attributes and outcomes. Direct discrimination was modeled as the causal effect occurring along the direct path from sensitive attributes to the outcome. On the other hand, indirect discrimination was represented by the causal effects along other paths involving non-sensitive features.

Li and Xu \cite{iccvLiX21} proposed a method to detect unknown biased attributes in a classifier that predicts a target attribute, such as gender, based on input images. The biased attribute in question is distinct from the target attribute. For instance, if a gender classifier exhibits varying predictions for female images based on their skin tones, then the skin tone attribute would be considered biased. To identify this unknown biased attribute, the authors optimized a hyperplane within the latent space of a generative model. By analyzing the transformation of synthesized counterfactual images generated by the model, human observers can interpret the semantic meaning of the biased attribute hyperplane. For example, images transitioning from ``light skin'' to ``dark skin'' indicate the presence of bias associated with skin color.

Black et al. \cite{fatBlackYF20} utilized optimal transport projections \cite{villani2009optimal} to map data instances from the unprivileged group to their counterparts in the privileged group. This allowed them to extract the positive flipset, which consists of unprivileged group members with favorable outcomes whose counterparts experienced unfavorable outcomes. Additionally, they computed the negative flipset, consisting of unprivileged group members with unfavorable outcomes whose counterparts experienced favorable outcomes. By analyzing the members of these flipsets, they determined which features contributed to inconsistent classifications.

\subsubsection{Detecting Label Bias} 
Label bias occurs when factors unrelated to the determination of labels influence the process that generates outcome labels \cite{nipsWickpT19}. ML models are often developed using data collected over an extended period. During the data collection process, labels are typically assigned by human annotators or algorithms, introducing the potential for human and algorithmic biases to be encoded into the labels.

To detect label bias, Chakraborty et al. \cite{sigsoftChakrabortyMM21,sigsoftChakrabortyM0M20,fairsslware} employed situation testing to identify biased data points and remove them from the training data. They divided the dataset into privileged and unprivileged groups based on the sensitive attribute. Two separate models were then trained on the data from each group. For each training instance, the predictions from both models were compared. If the two models produced divergent results, there was a probability that the label of that data point was biased.

Chen and Joo \cite{iccvChenJ21} utilized facial action units, objective indicators of fundamental muscle actions associated with different facial expressions, to detect label bias in widely-used datasets for facial expression recognition. They demonstrated that many expression datasets exhibited significant label bias between different gender groups, particularly concerning expressions of happiness and anger. Furthermore, they found that conventional fairness repair methods were unable to completely mitigate such biases in trained models.

\subsubsection{Detecting Selection Bias}
Selection bias occurs when the process of sampling training data introduces an unexpected correlation between sensitive attributes and the outcome \cite{nipsWickpT19}. For example, the Compas dataset \cite{compasdata}, widely studied in the fairness literature, has been shown to exhibit unintended correlations between race and recidivism \cite{nipsWickpT19}. This dataset was collected during a specific time period (2013 to 2014) and from a particular county in Florida, with its inherent policing patterns, making it susceptible to the introduction of unintentional correlations.

Researchers primarily employ distribution testing to detect selection bias in the data. Chen et al. \cite{zhenpengmaat22} tested whether the training data satisfied the ``We Are All Equal'' worldview, which assumes that there should be no statistical association between the outcome and the sensitive attribute. They specifically examined whether the favorable rates of privileged and unprivileged groups were equal. Chakraborty et al. \cite{sigsoftChakrabortyMM21,fairsslware} not only analyzed the disparity in favorable rates between privileged and unprivileged groups but also compared the numbers of data instances in the two groups.

K{\"{a}}rkk{\"{a}}inen and Joo \cite{KarkkainenJ21} detected bias in public face datasets, revealing a strong bias toward Caucasian faces while other racial groups (e.g., Latino) were significantly underrepresented. Such biases increase the risk of introducing fairness issues in facial analytic systems and limit their applicability.

Similarly, Torralba and Efros \cite{cvprTorralbaE11} investigated computer vision datasets to evaluate whether existing datasets genuinely represent unbiased representations of the real world. They assessed how well an object detector trained on one dataset generalized when tested on a representative set of other datasets.

Yang et al. \cite{aiesYangGFSY0NHJ22} collected perceived demographic attributes on a popular face detection dataset \cite{widerfacelink} and observed skewed demographic distributions. Face detectors trained on this dataset exhibited demographic bias, as measured by performance disparities among different groups.

Wang et al. \cite{eccvWangNR20} detected selection bias in visual datasets across three dimensions: object-based bias, gender-based bias, and geography-based bias. Object-based detection considered statistics related to object size, frequency, context, and diversity of object representation. Gender-based detection revealed the stereotypical portrayal of individuals of different genders. Geography-based detection focused on the representation of different geographic locations.

Mambreyan et al. \cite{datasetbiasicpr} analyzed datasets used for lie detection and discovered significant sex bias within them. Specifically, the percentage of instances labeled as lies for females was greater than that for males in the dataset. They further examined the effect of this bias on lie detection, training a classifier to predict the sex of the identity in a video and using sex as a proxy for lies (predicting lies for females and truth for males). This deception detector simulated a classifier that relied solely on selection bias. The results demonstrated that the performance of this biased classifier was comparable to the state-of-the-art, suggesting that recent techniques claiming near-perfect results may exploit selection bias.

\subsection{ML Program Testing}\label{programtesting}
An ML program includes various parts such as data processing, decision-making logic, and run-time configurations (e.g., ML hyper-parameters) \cite{sigsoftChenCLW0L20}. Each of these parts can potentially introduce a discordance between the existing fairness conditions and the desired ones for the final  ML software system. Testing the ML program can help identify fairness issues within its implementation.

\subsubsection{Testing Data Processing} 
In ML programs, it is common to include data processing scripts to manipulate and transform training data for downstream learning tasks. The processing can have a significant impact on the fairness of the software. 

Biswas and Rajan \cite{biswas2021fair} and Valentim et al. \cite{issreValentim0A19} have investigated the introduction of fairness bugs through data processing methods using causal reasoning. They systematically intervened in the development process of ML software by applying different commonly-used data processing methods while keeping other settings unchanged. Their findings indicate that certain pre-processing methods indeed introduce fairness bugs, while other methods may improve software fairness.

Caton et al. \cite{jairCatonMH22} have observed that real-world datasets often contain missing values, and one common approach to address this issue is to impute the missing values using various techniques during the data processing phase. To evaluate the impact of different imputation strategies on fairness outcomes, they conducted tests and examined the resulting fairness implications.

\subsubsection{Testing Hyper-parameters} 
The hyper-parameters specified in ML programs can affect fairness. To validate it, researchers explore whether different hyper-parameter settings lead to varying levels of software fairness. This testing process is treated as a search-based problem, aiming to discover optimal settings within the hyper-parameter space.

Chakraborty et al. \cite{corrabs190505786, sigsoftChakrabortyM0M20} proposed Fairway, a method that combines situation testing with multi-objective optimization. Since there is often a trade-off between fairness and ML performance (such as accuracy) \cite{CorbettDaviesP17}, Fairway employs sequential model-based optimization \cite{tseNair0MSA20} to search for hyper-parameters that maximize software fairness while minimizing any negative impact on other performance measures.

Similarly, Tizpaz-Niari et al. \cite{icse22fair1} considered both fairness and accuracy in their approach. They introduced Parfait-ML, which offers three dynamic search algorithms (independently random, black-box evolutionary, and gray-box evolutionary) to approximate the Pareto front of hyper-parameters that balance fairness and accuracy. Parfait-ML not only provides a statistical method to identify hyper-parameters that systematically influence fairness but also incorporates a fairness repair method to discover improved hyper-parameter configurations that simultaneously enhance fairness and accuracy.

Gohar et al. \cite{icse23ensemble} conducted fairness testing on hyper-parameters in ensemble learning. As ensemble hyper-parameters are more intricate due to their impact on how learners are combined within different ensemble categories, the researchers investigated the effects of ensemble hyper-parameters on fairness. They also presented how to design fair ensembles using ensemble hyper-parameters.

\subsubsection{Testing Fairness Repair Algorithms} 
As fairness becomes an increasingly important requirement for software systems, engineers may incorporate fairness repair algorithms (also known as bias mitigation algorithms) into their programs to ensure fairness. Researchers focus on testing whether these fairness repair algorithms effectively reduce fairness bugs without introducing side effects such as a decrease in accuracy.

Biswas and Rajan \cite{biswas2020machine} applied seven fairness repair algorithms to 40 top-rated ML models collected from a crowdsourced platform. They compared individual fairness, group fairness, and ML performance before and after applying these algorithms.

Qian et al. \cite{qian2021my} applied fairness repair techniques to five widely-adopted ML tasks and examined the variance of fairness and ML performance associated with these techniques. They investigated whether identical runs with a fixed seed produced different results. The findings indicated that most fairness repair techniques had undesirable impacts on the ML software, such as reducing accuracy, increasing fairness variance, or increasing accuracy variance.

Zhang and Sun \cite{junsunfse22} evaluated existing fairness repair techniques on DNNs and discovered that while these techniques improved fairness, they often resulted in a significant drop in accuracy. In some cases, fairness and accuracy were both worsened. They proposed an adaptive approach that selects the fairness repair method for a DNN based on causality analysis \cite{icseSun0PS22}.

Hort et al. \cite{sigsoftHortZSH21} introduced a benchmarking framework called Fairea. Prior work often measured the impacts of fairness repair algorithms on fairness and ML performance separately, making it unclear whether the improved fairness was solely due to the unavoidable loss in ML performance. Fairea addressed this issue by providing a unified baseline to evaluate and compare the fairness-performance trade-off of different repair methods.

Chen et al. \cite{chentosemstudy} utilized Fairea to conduct a large-scale, comprehensive empirical evaluation of 17 representative bias mitigation methods from both the ML and SE communities. They evaluated these methods across 12 ML performance metrics, 4 fairness metrics, and 24 types of fairness-performance trade-off measurements.

Hort and Sarro \cite{kbseHortS21} observed another side effect of fairness repair: it could lead to the loss of discriminatory behaviors of anti-protected attributes. Anti-protected attributes refer to attributes on which one might want the ML decision to depend (e.g., students with completed homework should receive higher grades).

Orgad et al. \cite{naaclOrgadGB22,orgad2022choose} evaluated fairness repair approaches for NLP models from two aspects: extrinsic bias (performance difference across different demographic groups) and intrinsic bias (bias in models' internal representations, e.g., sentence embeddings). They found that the two types of bias may not be correlated, and the choice of bias measurement and dataset can significantly affect the evaluation results.

\subsubsection{Testing Compression Algorithms} 
A computation-intensive DL model can be efficiently executed on PC platforms with GPU support, but it cannot be directly deployed and executed on platforms with limited computing power, such as mobile devices \cite{sigsoftChenCLW0L20}. To address this issue, model compression algorithms have been proposed to represent DL models in a smaller size with minimal impact on their performance \cite{icseChenYLCLWL21}. Common model compression techniques include quantization (representing weight values of DL models using smaller data types), pruning (eliminating redundant weights that contribute little to the model's behavior), and knowledge distillation (transferring knowledge from a large model to a smaller one) \cite{abs171009282}. The widespread adoption of model compression for DL models has motivated researchers to detect fairness bugs introduced by these algorithms.

Since model compression is often applied to large DL models, existing fairness testing of model compression algorithms typically focuses on complex NLP models \cite{corrabs210607849} and computer vision models \cite{abs201003058,corrabs220101709,corrabs210607849,eccv2022compression}. Hooker et al. \cite{abs201003058} demonstrated that pruning and quantization can amplify gender bias when classifying hair color in a computer vision dataset. Xu and Hu \cite{corrabs220108542} tested the effect of distillation and pruning on bias in generative language models and provided empirical evidence that distilled models exhibited less bias. Stoychev and Gunes \cite{corrabs220101709} detected fairness bugs introduced by different model compression algorithms in various facial expression recognition systems but did not observe consistent findings across different systems.

\subsection{Model Testing}\label{modeltesting}
Most existing fairness testing techniques primarily focus on the evaluation of individual ML models \cite{sigsoftGalhotraBM17,icseMingFan,kbseUdeshiAC18,sigsoftAggarwalLNDS19,trustcomXieW20,icseMingFan,icseZhangW0D0WDD20,isstaZhangZZ21,icse22fairNEURONFAIR}. These techniques can be directly applied to the final ML models obtained, using either a black-box or white-box approach. The distinction between white-box testing and black-box testing lies in the level of access to training data and internal knowledge of the ML models.

\subsubsection{Black-box Model Testing}
Black-box model testing is a technique used to detect fairness issues in ML models without relying on access to training data or knowledge of the internal model structure. This approach primarily relies on analyzing the behavior of the model based on the input space.

Fairness testing in the field typically relies on statistical measurements to identify fairness bugs in black-box models based on their prediction behaviors. For instance, Tramèr et al. \cite{TramerAGHHHJL17} conducted an analysis to detect fairness bugs by examining the associations between prediction outcomes and sensitive attributes. They aimed to uncover any potential biases or unfairness present in the model's predictions. Similarly, Bae \cite{bae2022discovering} compared the performance of pedestrian trajectory prediction models across different demographic groups, aiming to uncover variations or biases in their performance.

Additionally, fairness testing approaches often leverage metamorphic relations to detect fairness bugs by applying transformations to software inputs. These transformations aim to identify unexpected changes in the model's predictions. Many of these techniques employ black-box testing methodologies. For example, the Themis tool \cite{sigsoftGalhotraBM17,sigsoftAngellJBM18} generates random test inputs and checks if the software system produces consistent outputs for individuals who differ only in sensitive attribute values. Similarly, Aequitas \cite{kbseUdeshiAC18} and ExpGA \cite{icseMingFan} search the input space of the software for discriminatory instances that reveal unfair predictions.

Black-box testing is commonly used to detect fairness bugs in complex software systems, including NLP, computer vision, and ranking systems, where the internal workings of the system are not fully visible to the testers.

Researchers have developed various text templates to uncover fairness bugs in different NLP systems, such as sentiment analysis \cite{starsemKiritchenkoM18,9653830}, machine translation \cite{aclWangRC22}, text generation \cite{emnlpShengCNP19,emnlpHuangZJSWRMYK20,fatDhamalaSKKPCG21}, natural language inference \cite{sharma2021evaluating}, named entity recognition \cite{htMehrabiGMPG20}, and conversational systems \cite{biasaskerfse23}. Further details regarding these studies can be found in Section \ref{domaingeneration}.

For computer vision systems, state-of-the-art fairness techniques \cite{GoodfellowPMXWOCB14,corrabs190606439,KarrasALL18,corrabs200510430,corabs211108856,denton2019image,wacvDashB022} often utilize GAN-based algorithms to generate images that differ in sensitive attributes. These techniques then check if the computer vision systems make different decisions for equivalent image mutants. More information about these techniques is available in Section \ref{domaingeneration}.

Ranking systems, being predominantly black-box, are also tested in a black-box manner \cite{singh2018fairness,yang2017measuring,kuhlman2019fare,NarasimhanCGW20,BeutelCDQWWHZHC19}. For instance, researchers have measured whether different demographic groups have proportional representation in top-ranking positions based on the system's ranking outputs. FairRec \cite{isstaFairRec} assesses recommendation differences across demographic groups using statistical measurements such as recommendation performance, alignment of recommended item popularity with user preferences, and diversity in recommendations.

Some black-box testing techniques approximate the behavior of the black-box software using a white-box model, allowing the application of white-box testing techniques. For example, Aggarwal \cite{sigsoftAggarwalLNDS19} approximated the decision-making process of the black-box ML software using a decision tree constructed through a local model explainer. They then employed symbolic execution-based test input generation to discover discriminatory inputs. Sharma and Wehrheim \cite{ptsSharmaW20} first approximated the black-box software with a white-box model based on its behaviors. They subsequently developed a property-based testing mechanism for fairness checking, where specific fairness requirements can be specified using an assume-assert construct. Test cases were automatically generated to attempt to violate the specified fairness property.

\subsubsection{White-box Model Testing}
White-box model testing aims to identify fairness bugs by examining either the training data or the internal structure and information of the ML model that is accessible to test engineers.

Some approaches leverage \emph{training data} to uncover unfair predictions, without accessing the internal of ML models. Chakraborty et al. \cite{kbseChakrabortyPM20} proposed an explanation method based on k-nearest neighbors to detect bias in ML software predictions. They identified instances predicted unfavorably and examined their k-nearest neighbors with favorable labels from the training data. By comparing the distribution of these neighbors with the test instance, they determined and explained the presence of bias. 

Zhao et al. \cite{emnlpZhaoWYOC17} used fairness measurements from training data as a baseline to identify bias amplification. They annotated unlabeled data using the ML model and compared fairness measurements between the training data and the annotated dataset to reveal bias amplification. 

Wang and Russakovsky \cite{icmlWangR21} highlighted issues with Zhao et al.'s bias amplification measurement and proposed a new, decoupled metric that considers varying base rates of sensitive attributes.

Cabrera et al. \cite{cabrera2019fairvis} developed FAIRVIS, a testing tool for subgroup fairness. It efficiently searches for potential issues among numerous subgroups, clustering the training dataset to identify statistically similar subgroups, and calculating group fairness metrics using an entropy-based approach. FAIRVIS presents generated subgroups sorted by their group fairness metrics.

Patel et al. \cite{icstPatelCLKK22} utilized combinatorial t-way testing~\cite{kuhn2013introduction} for fairness testing. This coverage-based data sampling method generates diverse datasets by applying logical constraints. They created an input parameter model from the training data and used it to generate a t-way test set. Discriminatory instances were identified by mutating protected attributes in each test.

Several techniques \cite{icseZhangW0D0WDD20,9506918,guanhongfse22,icse22fairNEURONFAIR,isstaZhangZZ21,corabs211108856} utilize \emph{gradient information}, which represents the direction of steepest ascent in the loss function, to generate test inputs for fairness testing. ADF \cite{icseZhangW0D0WDD20,9506918} focuses on discriminatory instances near the decision boundary of DNNs and employs gradients to guide the search for neighboring test inputs. EIDIG \cite{isstaZhangZZ21} reduces gradient calculations to accelerate the search process.

Researchers have also developed various methods to detect and analyze \emph{neurons} responsible for unfair outcomes in emerging deep learning (DL) models. NeuronFair \cite{icse22fairNEURONFAIR} utilizes neuron analysis to identify biased neurons contributing to unfairness, and generates discriminatory instances to amplify the activation differences of these biased neurons. It demonstrates strong interpretability, generation effectiveness, and data generalization.

DeepFAIT \cite{corabs211108856} employs significance testing to identify fairness-related neurons by analyzing the activation differences between privileged and unprivileged groups.

Vig et al. \cite{nipsVigGBQNSS20} apply Causal Mediation Analysis (CMA) to identify causally implicated parts, such as neurons or attention heads, in the unfair predictions of a DNN model. CMA measures the direct and indirect effects of targeted neurons on the final unfair predictions, considering each neuron as an intermediate mediator.

Tian et al. \cite{icseTianZOKR20} introduce a novel statistical measurement using neuron activation in DNNs. They compute neuron activation vectors for each label class based on test inputs and calculate the distance between these vectors to assess fairness. Dissimilar distances with respect to a third class indicate the presence of fairness bugs in the tested DNN.

Gao et al. \cite{icse22fair3} propose FairNeuron for DNNs, which employs neuron slicing to identify conflict \emph{paths} containing neurons that rely on sensitive attributes for predictions. Biased instances triggering the selection of sensitive attributes are identified using these paths, and the model is retrained through selective training. FairNeuron ensures that the conflict paths learn all important features for prediction instead of biased ones for the identified biased instances, while retaining the original training approach for other instances.

Additionally, Zhang et al. \cite{JiangZhangcorr} developed a method to identify and rectify fairness-related \emph{paths} in decision tree and random forest models. They employed a MaxSMT solver to determine the paths that could be altered while satisfying fairness and semantic difference constraints. The identified paths were refined by modifying the leaf labels, resulting in a repaired fair model.

\section{Research Trends and Distributions}\label{distri}
In Figure \ref{fig:trend}, we have shown that fairness testing is experiencing a dramatic increase in the number of publications.
This section further analyzes the research trends and distributions of fairness testing.

\begin{figure}[t]
    \centering
      \includegraphics[width=0.7\linewidth]{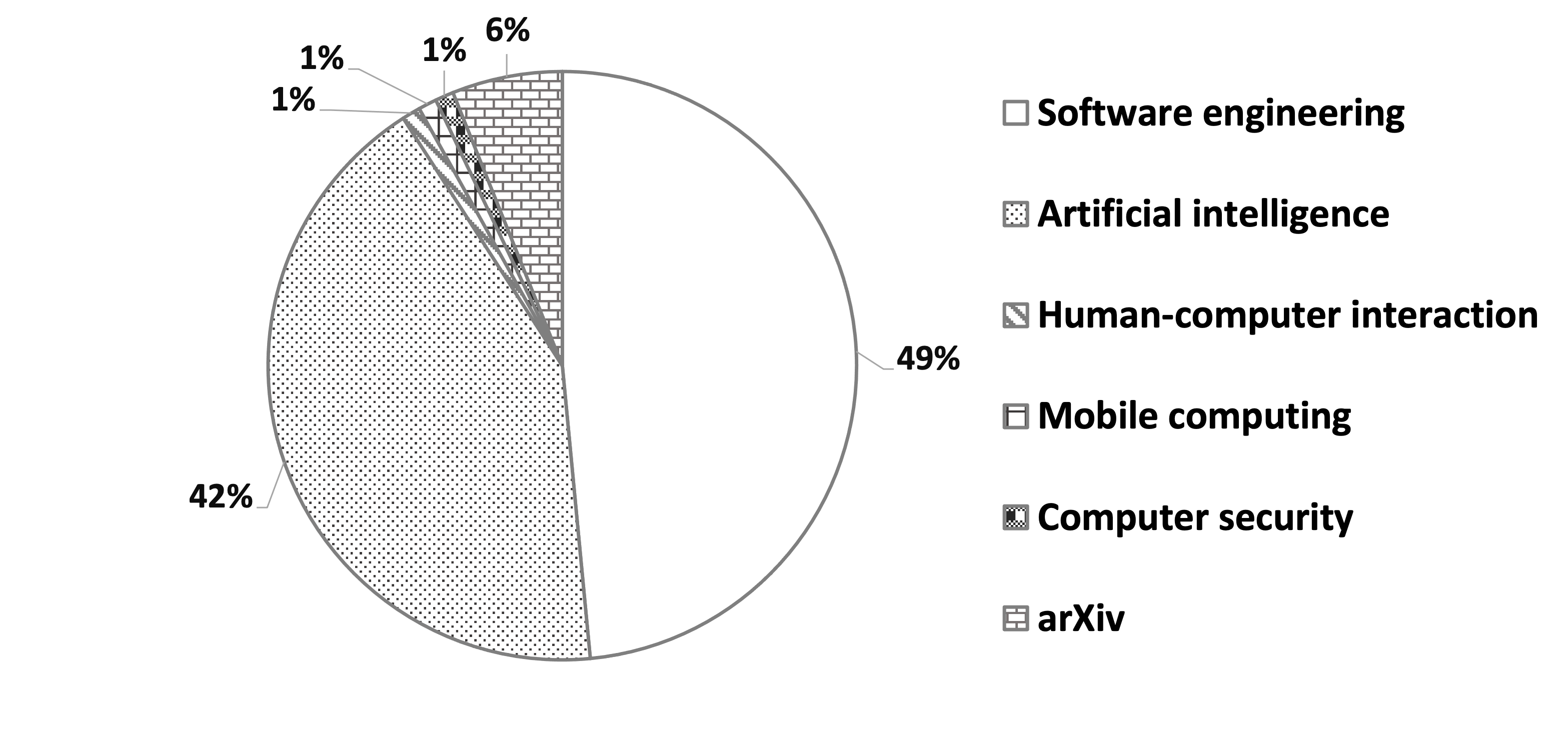}
  \caption{Research distribution among different venues.}
  \label{fig:venue} 
\end{figure}

\subsection{Research Venues}
We first describe research trends in terms of the research communities engaging in fairness testing.
Since 2017, an increasing number of research communities have dedicated their efforts to studying fairness testing. Notable contributions include:
\begin{itemize}
\item Galhotra et al. \cite{sigsoftGalhotraBM17} introduced the first fairness testing approach for ML software in the SE community, receiving the Distinguished Paper Award at ESEC/FSE 2017.
\item Zhao et al. \cite{emnlpZhaoWYOC17} detected bias in datasets and ML models for visual recognition tasks, earning the Best Paper Award at EMNLP 2017.
\item D{\'{\i}}az et al. \cite{chiDiazJLPG18} identified age-related bias in sentiment analysis systems, receiving the Best Paper Award at CHI 2018.
\item Ribeiro et al. \cite{aclRibeiroWGS20} proposed CheckList, a task-agnostic methodology for testing NLP models, including fairness testing, which received the Best Paper Award at ACL 2020.
\item Zhang et al. \cite{icseZhangW0D0WDD20} proposed a search-based discriminatory instance generation approach for DNNs, which received the Distinguished Paper Award at ICSE 2020 and was selected for SIGSOFT Research Highlights.
\item Chakraborty et al. \cite{sigsoftChakrabortyMM21} addressed selection bias and label bias in training data, and were honored with the Distinguished Paper Award at ESEC/FSE 2021.
\end{itemize}

These six best paper awards, as well as being a credit to their authors, also demonstrate both the significant level of interest and the high quality of research on fairness in three different research communities:
\begin{itemize}
\item Software Engineering (ICSE and ESEC/FSE)
\item Natural Language Processing (EMNLP and ACL)
\item Human-Computer Interaction (CHI)
\end{itemize}

Figure \ref{fig:venue} illustrates the distribution of the collected papers across various research venues. The majority of fairness testing papers (49\%) are published in software engineering venues, including ICSE, ESEC/FSE, ASE, ISSTA, TSE, and TOSEM. Artificial intelligence venues, such as ICML, NeurIPS, IJCAI, ACL, EMNLP, CVPR, ECCV, and KDD, account for 42\% of the fairness testing papers.

Furthermore, our survey reveals that fairness testing is gaining traction in other research communities, such as computer security, human-computer interaction, and mobile computing communities. This highlights the broad audience and significance of our survey across multiple disciplines.

\subsection{Machine Learning Categories}
In this section, we explore the research trend of fairness testing across different ML categories. Following previous work \cite{tseZhangHML22}, we categorize the gathered papers into two groups: those that concentrate on DL software and those that address general ML software.

Out of the total number of papers, 50 papers (50\%) focus on conducting fairness testing for DL software, 41 papers (41\%) specifically target general ML software, and 8 papers (8\%) consider both traditional ML software and DL software. The significant volume of publications on fairness testing for DL software can be attributed to several factors. On one hand, DL has gained widespread adoption and is being utilized in a diverse range of software applications, generating significant interest from the research community. On the other hand, compared to traditional ML algorithms such as regression and decision trees, DL models are less interpretable \cite{sigiteBaranyiNM20}, making it more challenging to directly reason about fairness.

To gain further insights, we analyze the publication trends for both categories over the years. Figure \ref{fig:ml_trend} depicts the number of papers focusing on fairness testing in general ML and DL per year. Our analysis reveals a clear shift in research focus, with a transition from testing general ML software to testing DL software. Prior to 2019, fairness testing research primarily concentrated on general ML. However, since 2019, the number of papers specifically addressing DL has experienced a notable surge, surpassing the publications on general ML.

\begin{figure}[t]
    \centering
      \includegraphics[width=0.6\linewidth]{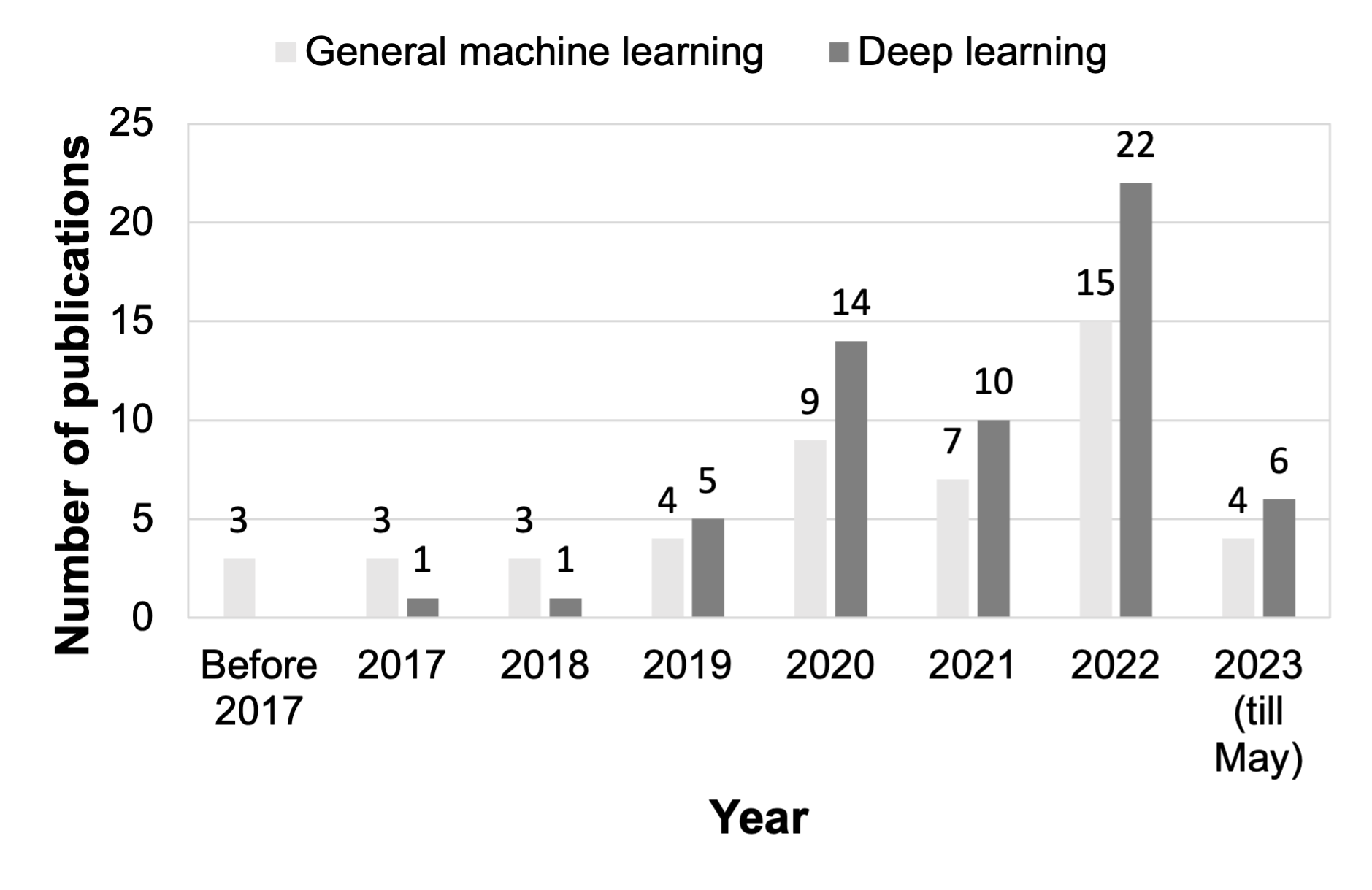}
  \caption{Number of fairness testing papers on general machine learning and deep learning per year.}
  \label{fig:ml_trend} 
\end{figure}

\subsection{Data Types}\label{datatypesection}
In this section, we investigate  the research trends of fairness testing in applications that involve different types of data.

Out of the total number of papers (\num) that we have collected, 5 of them consider more than one data type. We count each of these papers for each data type they examine, allowing us to analyze the distribution across various data types. The findings are illustrated in Figure \ref{fig:datatype}.

\begin{figure}[t]
    \centering
      \includegraphics[width=0.6\linewidth]{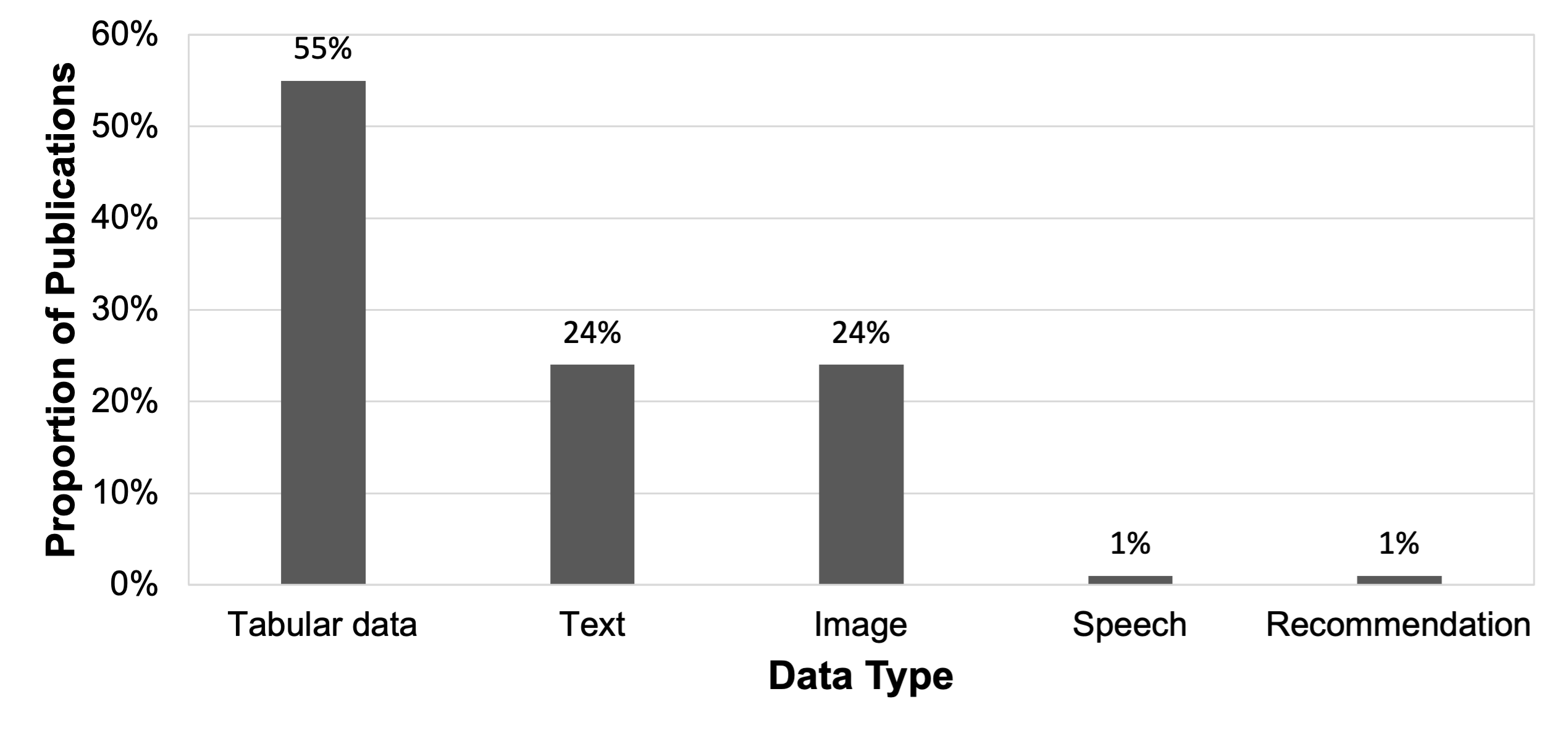}
  \caption{Distribution of different data types in fairness testing papers.}
  \label{fig:datatype} 
\end{figure}

\begin{figure}[t]
    \centering
      \includegraphics[width=0.6\linewidth]{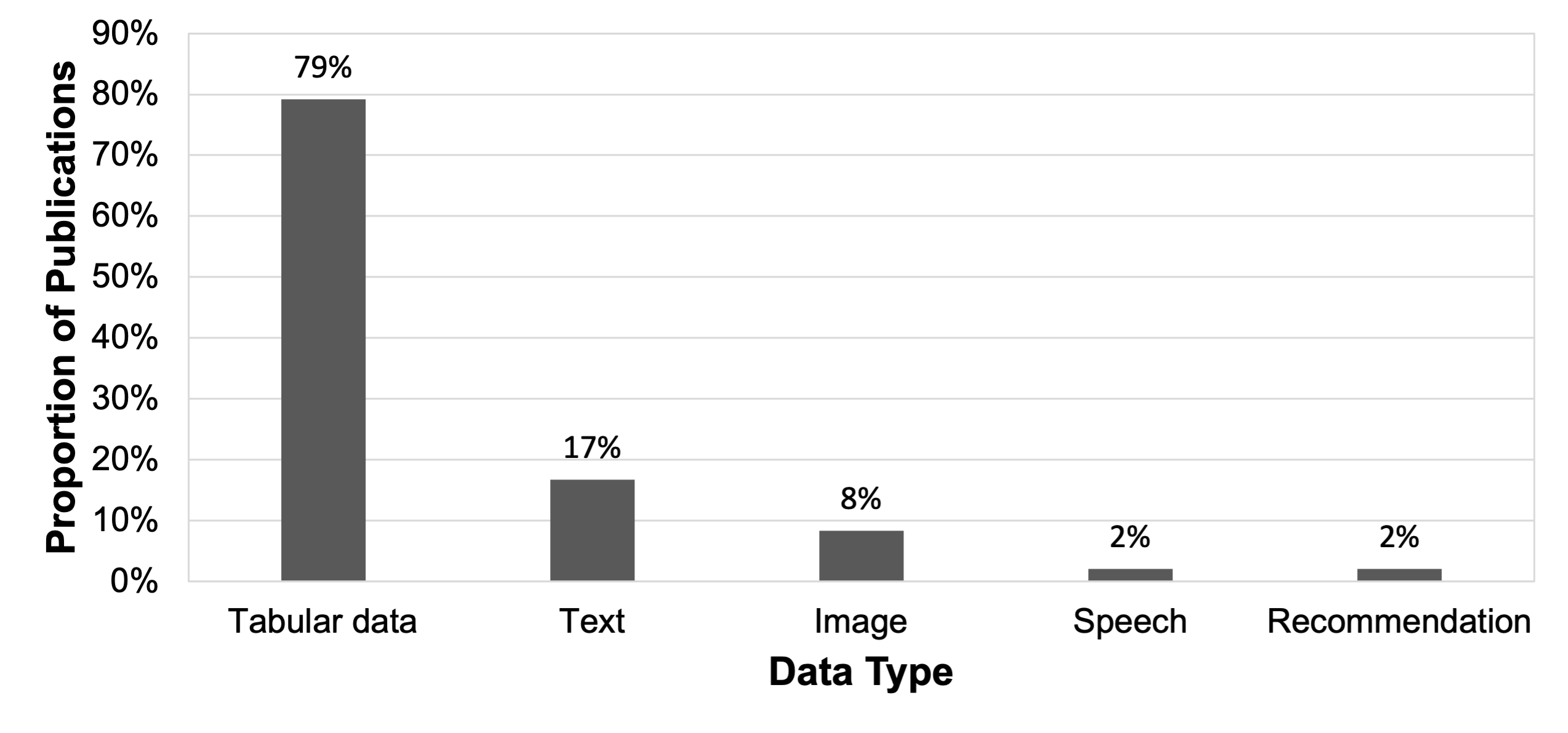}
  \caption{Distribution of different data types in fairness testing papers published in SE venues.}
  \label{fig:sedatatype} 
\end{figure}

Our analysis indicates that among the publications we have collected, a significant portion focuses on testing software applications that utilize tabular data as inputs, accounting for 55\% of the total. Furthermore, approximately 24\% of publications address fairness testing in applications involving text inputs, while another 24\% specifically tackle fairness issues in applications utilizing image inputs.

It is important to note that fairness testing for other data types, such as speech and recommendation systems, has not yet received extensive investigation, representing only a small percentage of the publications, approximately 1\% each. These data types remain relatively underexplored in the context of fairness testing, emphasizing the need for further research and attention to ensure fairness across a wider range of data-driven applications.


We also plot the data type distribution for SE publications. Figure \ref{fig:sedatatype} shows the results. In comparison to the broader research landscape, the SE community predominantly focuses on applications involving tabular data, representing a substantial majority of fairness testing publications in SE venues, amounting to 79\%. In contrast, publications addressing text- or image-based problems account for only 17\% and 8\%, respectively. These figures are significantly lower than the average distribution across all data types.

\subsection{Fairness Categories}

\begin{figure}[t]
    \centering
      \includegraphics[width=0.6\linewidth]{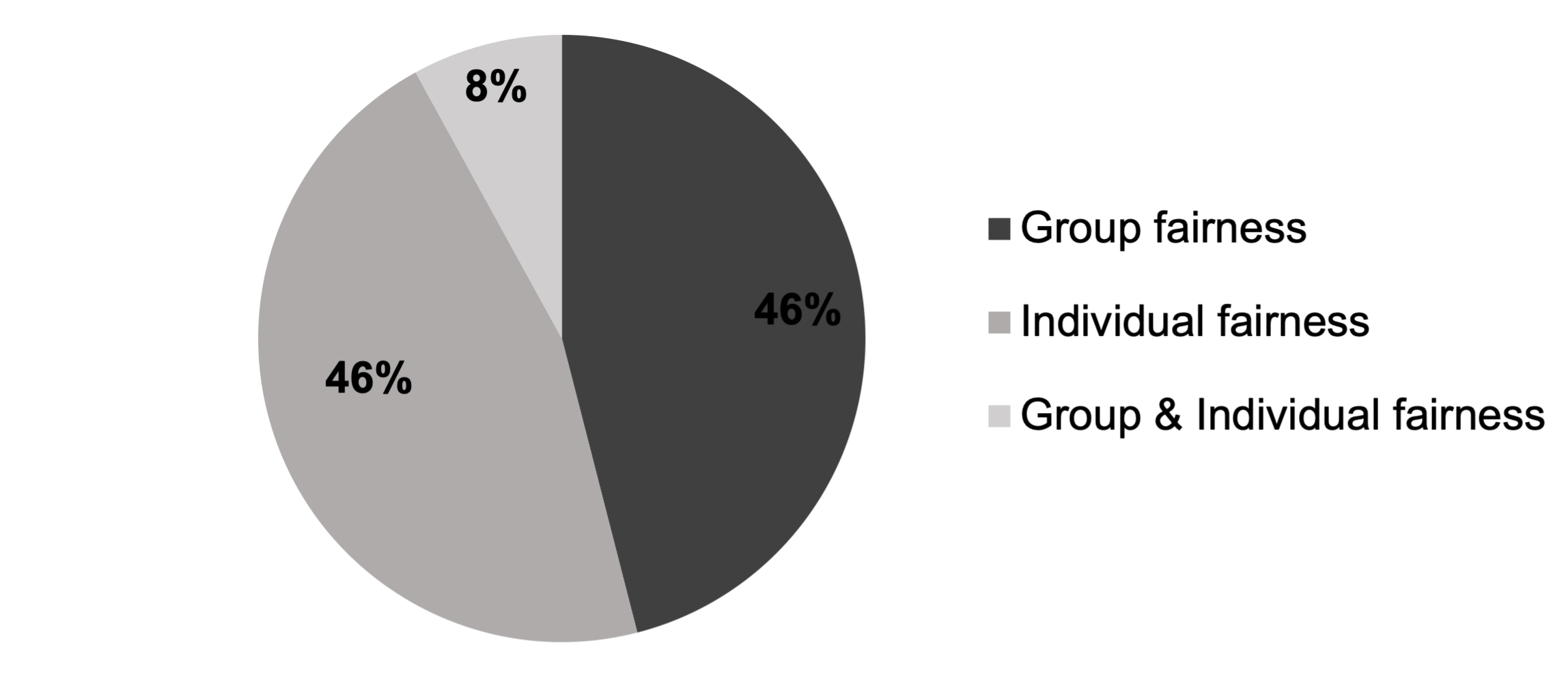}
  \caption{Distribution of different fairness categories.}
  \label{fig:property} 
\end{figure}

Figure \ref{fig:property} provides insights into the distribution of different fairness categories within the fairness testing literature. Our analysis reveals that comparable research efforts have been dedicated to exploring various aspects of fairness testing.

Specifically, we observe that 46\% of fairness testing papers focus on group fairness, addressing issues related to fairness among different groups. Another 46\% of papers concentrate on individual fairness, investigating fairness concerns at the individual level. The remaining 8\% of papers examine both group fairness and individual fairness, recognizing the importance of considering both perspectives in fairness testing research.

The finding may initially appear contradictory when comparing it with Table \ref{testg_summary}, where we observe that nearly all test generation techniques are proposed for individual fairness. This discrepancy can be attributed to the distinct characteristics of individual fairness and group fairness.

When examining group fairness, researchers can leverage real-world data to construct test inputs that capture fairness considerations among different groups. This availability of data facilitates the exploration of group fairness in testing scenarios. On the other hand, for individual fairness, it becomes challenging for researchers to identify pairs of instances that satisfy the specific input requirements associated with individual fairness. For instance, it is not always straightforward to find two individuals who differ solely in a sensitive protected attribute, making test input generation more focused on individual fairness.

Despite these differences in test input generation, it is important to note that fairness testing studies overall investigate group fairness and individual fairness at a similar level. The distribution of research efforts aims to address both perspectives, acknowledging the significance of both group and individual fairness in testing methodologies. 

\subsection{Testing Manners}
We further categorize existing fairness testing techniques based on the software testing approach employed, distinguishing between white-box and black-box methods. It is worth noting that although Tizpaz-Niari et al. \cite{icse22fair1} proposed black-box and gray-box techniques for testing hyper-parameters, we consider their approach as white-box because it requires access to the training data for model training and fairness evaluation.

In our analysis, we find that 53\% of the \num papers employ black-box testing techniques, while 47\% adopt white-box testing methods. This observed gap between black-box and white-box testing proportions is reasonable.

Compared to black-box testing, white-box testing necessitates access to either training data or the internal workings of software systems. However, fairness testing often applies to systems that are human-related and have social significance, making it challenging to disclose internal information to the public due to privacy concerns or legal policies. Therefore, it is reasonable that more research studies focus on black-box testing approaches, which do not require detailed knowledge of the internal system mechanisms.

\subsection{Tasks and Sensitive Attributes}
In the collected papers, researchers typically present their techniques and then employ datasets featuring specified tasks and sensitive attributes, for technique evaluation. In this section, we provide a summary of prevalent tasks and sensitive attributes commonly considered in the fairness testing literature.

We present in Table \ref{comtask_info} commonly studied tasks in fairness testing literature. As detailed in Section~\ref{datatypesection}, existing fairness testing techniques span various data types, enabling support for diverse tasks. Notably, income prediction, credit risk prediction, and recidivism/crime prediction emerge as the three most widely studied. These tasks also feature as the most widely-explored ones in the bias mitigation (i.e., fairness improvement) literature, as indicated by a recent survey \cite{maxfairnesssurvey}.

We list in Table \ref{sena_info} commonly studied sensitive attributes in the fairness testing literature. Overall, the literature covers a wide range of sensitive attributes, including sex/gender, race/ethnicity, age, country, occupation, religion, and sexual orientation. Notably, the top three sensitive attributes are sex/gender, race/ethnicity, and age, which are consistent with previous findings observed in the general software fairness literature \cite{corrabs220508809}. They are considered in 89\%, 57\%, and 41\% of the collected papers, respectively.

\begin{table}[!tp]
\footnotesize
\centering
\caption{Commonly-studied tasks in the fairness testing literature.}
\label{comtask_info}
\begin{tabular}{lr}
\hline
Task & \#Publications\\
\hline
Income prediction & 49\\
Credit risk prediction & 33 \\
Recidivism/crime prediction & 26\\
Deposit subscription prediction & 24 \\
Face recognition/facial analysis & 16\\
Healthcare prediction & 10 \\
Conversational AI/language generation & 9 \\
Sentiment analysis & 8\\
Heart heath prediction & 7 \\
Image tagging/classification & 7\\
Exam performance/grade prediction & 5\\
Coreference resolution & 3\\
Machine translation & 3\\
Individual survival prediction & 3\\
Recommender systems & 2\\
Pedestrian trajectory prediction & 1\\ 
Admission prediction & 1\\
Loan application & 1\\
Hiring & 1\\
Pricing & 1\\
Fraud detection & 1\\
Car rental & 1\\
Execution prediction & 1\\
Toxicity detection & 1\\
Name entity recognition & 1\\
Speech recognition & 1 \\
COVID-19 new case prediction & 1\\
Emergency department wait-time prediction & 1\\
\hline
\end{tabular}
\end{table}

\begin{table}[!tp]
\footnotesize
\centering
\caption{Commonly-studied sensitive attributes in the fairness testing literature.}
\label{sena_info}
\begin{tabular}{lr}
\hline
Sensitive attribute & \#Publications\\
\hline
Sex/gender & 89 \\
Race/ethnicity/skin tone & 57 \\
Age & 41\\
Country/nationality/geography & 8\\
Occupation/profession & 6\\
Religion & 3\\
Political ideology & 2 \\
Sexual orientation & 2\\
Person name & 2\\
Ability & 2\\
Body & 2\\
Character & 2\\
Culture & 2\\
Social status & 2  \\
Marital status & 1\\
Hair color & 1\\
Victim & 1\\
\hline
\end{tabular}
\end{table}

\section{Datasets and Tools}\label{data_Tool}
Based on the collected papers, this section summarizes the public datasets and open-source tools for fairness testing to provide a quick navigation for researchers and practitioners.


\begin{table*}[!tp]
\scriptsize
\centering
\caption{Public datasets for fairness testing.}
\label{dataset_info}
\begin{tabular}{llllll}
\hline
Dataset & Description & Data Type & Size & Sensitive Attribute(s) & URL\\
\hline
Adult Income & Income prediction & Tabular & 48,842 & sex, race & \cite{adultdata} \\
Census-Income (KDD) & Income prediction & Tabular & 299,285 & sex, race & \cite{incomekdddata}\\
Compas & Recidivism prediction & Tabular & 6,167 & sex, race  & \cite{compasdata} \\
German Credit & Credit risk prediction &Tabular & 1,000& sex & \cite{germandata}\\
Default Credit & Credit risk prediction &Tabular & 30,000 & sex & \cite{DefaultCreditdata}\\
Home Credit & Credit risk prediction &Tabular & 37,511 & sex & \cite{HomeCreditdata}\\
Bank Marketing &  Deposit subscription prediction & Tabular &30,488 & age & \cite{bankdata}\\
Mep15 & Health care needs prediction & Tabular &15,830 & race & \cite{mepdata}\\
Mep16 & Health care needs prediction & Tabular & 15,675 & race & \cite{mep16data} \\
Dutch Census & Income prediction & Tabular & 60,421 & gender & \cite{Dutchdata} \\
Heart Health & Heart health prediction & Tabular &303 & age & \cite{hearthealthdata}\\
Arrhythmia & Cardiac arrhythmia prediction& Tabular & 452 & sex & \cite{Arrhythmiadata}\\
Student Performance & Final year grade prediction &Tabular & 649 & sex & \cite{studentthdata}\\
UFRGS & GPA prediction & Tabular & 43,302 & gender, race & \cite{UFRGSdata} \\
Law School & Exam performance prediction & Tabular &  20,000 & gender, race & \cite{lawschooldata}\\
Titanic & Individual survival prediction & Tabular &891 & sex & \cite{Titanicdata}\\
Communities and Crime & Crime prediction &Tabular &2,215 & race, age & \cite{Crimedata} \\
Diabetes & Readmission prediction & Tabular &100,000 & race & \cite{Diabetesdata} \\
Heritage Health & Staying in the hospital or not &Tabular & 147,473 & age & \cite{Heritagedata}\\
Fraud  &  Fraud detection &Tabular & 1,100 & age & \cite{Frauddata} \\
US Executions & Execution prediction & Tabular &1,437 & sex, race & \cite{Executiondata}\\
LFW & Face recognition & Image &10,000 & sex, race & \cite{lfwlink} \\
BUPT-Transferface & Face recognition & Image & 1,300,000 & race & \cite{Transferfacedata} \\
VGGFace & Face recognition & Image & 2,600,000 & gender, race & \cite{VGGFacedata} \\
COCO-gender & Object detection & Image & 66 objects & gender & \cite{cocogender}\\
CelebA & Facial analysis & Image &202,599 & gender & \cite{CelebAdata}\\
RAF-DB & Facial analysis & Image & 15,339 & gender, race, age & \cite{raflink}\\
PBB & Facial analysis & Image & 1,270 & gender, skin type & \cite{pbbdataset} \\
FairFace & Facial analysis & Image & 108,501 & gender, race, age & \cite{FairFacedata}\\
WinoST & Speech translation & Audio & 3,888 & gender & \cite{WinoSTlink} \\
AequeVox & Speech recognition & Audio & 68 & gender, accent & \cite{AequeVoxdata}\\
WinoBias & Coreference resolution & Text& 3,160 & gender & \cite{WinoBiasdata} \\
Winogender &  Coreference resolution & Text& 720 & gender & \cite{Winogenderdata} \\
IMDB & Sentiment analysis & Text & 50,000 & gender, country, occupation & \cite{IMDBdata} \\
Twitter Sentiment140 & Sentiment analysis & Text & 1.6 million & gender, country, occupation & \cite{Sentiment140data} \\
SST & Sentiment analysis & Text & 11,855 & gender, country & \cite{sstdataset}\\
EEC & Sentiment analysis & Text &8,640 & sex, race & \cite{EECdata} \\
Large Movie Review & Sentiment analysis & Text & 50,000 & gender & \cite{largemovie_data}\\
Wikipedia comments & Toxicity detection & Text & 127,000 & religion, country, ethnic,  race & \cite{Wikipediacomment_data} \\
Jigsaw Comments & Toxicity classification & Text & 313,000 & religion, country, ethnic, race & \cite{Jigsaw_data}\\
nlg-bias & Language generation & Text & 360 & gender, race, sexual orientation & \cite{nlgbias_data} \\
BOLD & Language generation & Text & 23,679  & \makecell[l]{profession, gender, race,\\religion,  political ideology} & \cite{bolddata}\\
HOLISTICBIAS & Language generation & Text & 459,758 & 13 attributes & \cite{HOLISTICBIASdata}\\
DialogueFairness & Conversational AI & Text & 300,000 & gender & \cite{DialogueFairnesslink}\\
BiasAsker & Conversational AI & Text & 8,110 & \makecell[l]{ability, age, body character,\\ culture, gender, profession,\\ race, religion, social victim} & \cite{biasaskerlink}\\
MovieLens & Recommendation & Movie ratings & 1 million & gender, age, occupation & \cite{movielenslink} \\
LFM360K & Recommendation & \makecell[l]{Music listening\\history} & 17 million & \makecell[l]{gender, age,\\nationality} & \cite{LFM360Klink}\\
BlackFriday & Recommendation & Purchase history & 550,068 & \makecell[l]{gender, age, occupation,\\ city category, years of\\ stay in the current city,\\ marital status} & \cite{BlackFridaylink}\\
\hline
\end{tabular}
\end{table*}

\subsection{Public Datasets}\label{publicdata}
This section lists the public datasets in the literature based on our collected papers.
Table \ref{dataset_info} provides detailed information about these datasets, including their sizes, data types, sensitive attributes, usage scenarios, and access links. Many of these datasets comprise tabular data, making them suitable for traditional ML classifiers.

In recent years, there has been a surge in the availability of text and image datasets, driven by the growing popularity of natural language processing and computer vision. These datasets are often sourced from social media platforms like Twitter. It is worth noting that certain datasets come with specific usage constraints that researchers must consider when utilizing them. For instance, the well-known image dataset CelebA \cite{CelebAdata} is restricted to non-commercial research purposes only.

For a comprehensive overview of the existing fairness datasets, we recommend referring to the works of Le Quy et al. \cite{corrabs211000530} and Fabris et al. \cite{corrabs220201711}. Le Quy et al. \cite{corrabs211000530} surveyed tabular datasets specifically for fairness research, while Fabris et al. \cite{corrabs220201711} expanded the survey to include unstructured data such as text and images. Their surveys encompass datasets from diverse domains, including social sciences, computer vision, health, economics, business, and linguistics.

\begin{table*}[!tp]
\tiny
\centering
\caption{Open-source tools for fairness testing.}
\label{tools_info}
\begin{tabular}{lllll}
\hline
Tool [ref] & Application & Component & Description & URL\\
\hline
FairTest \cite{TramerAGHHHJL17} & General ML & Model & Analyzing associations between outcomes and sensitive attributes & \cite{FairTestlink}\\
Themis \cite{sigsoftGalhotraBM17} & \makecell[l]{Classification\\(Tabular data)}  & Model & Black-box random discriminatory instance generation  & \cite{Themislink} \\
Aequitas \cite{kbseUdeshiAC18} & \makecell[l]{Classification\\(Tabular data)} & Model & Automated directed fairness testing & \cite{Aequitaslink}\\
ExpGA \cite{icseMingFan} & \makecell[l]{Classification\\ (Tabular data, text)} & Model & Explanation-guided fairness testing through genetic algorithm & \cite{ExpGAlink} \\
fairCheck \cite{ptsSharmaW20} & \makecell[l]{Classification\\ (Tabular data)} & Model & Verification-based discriminatory instance generation & \cite{fairChecklink}\\
MLCheck \cite{sharma2021mlcheck} & \makecell[l]{Classification\\ (Tabular data)} & Model & Property-driven testing of ML  models & \cite{mlchecklink} \\
LTDD \cite{icse22fair2} & \makecell[l]{Classification\\ (Tabular data)} & Data & Detecting which data features and which parts of them are biased & \cite{LTDDlink}\\
Fair-SMOTE \cite{sigsoftChakrabortyMM21}& \makecell[l]{Classification\\ (Tabular data)} & Model & Detecting biased data labels and data distributions & \cite{SMOTElink}\\
FairMask \cite{peng2021xfair}& \makecell[l]{Classification\\ (Tabular data)} & Data  & Extrapolation of correlations among data features that might cause bias & \cite{xFAIRlink}\\
Fairway \cite{sigsoftChakrabortyM0M20} & \makecell[l]{Classification\\ (Tabular data)} & \makecell[l]{Data, ML\\ program} & Detecting biased data labels and optimal hyper-parameters for fairness& \cite{Fairwaylink}\\
Parfait-ML \cite{icse22fair1} & \makecell[l]{Classification\\ (Tabular data)} & ML program  & Searching for hyper-parameters optimal to ML fairness   & \cite{Parfaitlink}\\
Fairea \cite{sigsoftHortZSH21} & \makecell[l]{Classification\\ (Tabular data)} & \makecell[l]{ML program,\\ model} & A unified benchmark for evaluating fairness repair algorithms & \cite{Fairealink}\\
IBM AIF360 \cite{corrbs181001943} & \makecell[l]{Classification\\ (Tabular data)} & \makecell[l]{Data, ML \\program, model}  & Examining and mitigating bias in ML software & \cite{AIF360link}\\
I\&D \cite{corrabs220908321} & \makecell[l]{Classification\\ (Tabular data)}  &  Model & Improving initial individual discriminatory instances generation & \cite{idlink} \\
scikit-fairness \cite{skfairnesslink}  & \makecell[l]{Classification\\ (Tabular data)} & \makecell[l]{Data, ML\\ program, model} & Examining and mitigating bias in ML software & \cite{skfairnesslink} \\
LiFT \cite{cikmVasudevanK20} & \makecell[l]{Classification\\ (Tabular data)} & \makecell[l]{Data, ML\\ program, model}  & Examining and mitigating bias in ML software & \cite{liftlink} \\
FairVis \cite{cabrera2019fairvis} & \makecell[l]{Classification\\ (Tabular data)}  & Model & Visual analytics for discovering intersectional bias in ML software & \cite{fairvislink} \\
BiasAmp \cite{icmlWangR21} & Image classifier  & Model & Analyzing whether ML exacerbates bias from the training data & \cite{biasamplink} \\
MAAT \cite{zhenpengmaat22} & \makecell[l]{Classification\\ (Tabular data)}  & Data  & Detecting selection bias and improving fairness-performance trade-off & \cite{MAATlink} \\
FairEnsembles \cite{icse23ensemble} & \makecell[l]{Classification\\ (Tabular data)}  & ML program  & Analyzing fairness and its composition in ensemble ML & \cite{fensemblelink}\\
FairRepair \cite{JiangZhangcorr} & \makecell[l]{Tree-based classification\\ (Tabular data)}  & Model & Fairness testing and repair for tree-based models& \cite{FairRepairlink} \\
SBFT \cite{esefairnessserart} & \makecell[l]{Regression\\ (Tabular data)}  & Model & Search-based fairness testing for regression-based ML systems & \cite{SBFTlink}\\
ADF \cite{icseZhangW0D0WDD20} & \makecell[l]{DL-based classification\\ (Tabular data)}  & Model & White-box fairness testing through adversarial sampling & \cite{ADFlink} \\
EIDIG \cite{isstaZhangZZ21} & \makecell[l]{DL-based classification\\ (tabular data)} & Model & White-box fairness testing through gradient search & \cite{EIDIGlink} \\
NeuronFair \cite{icse22fairNEURONFAIR} & \makecell[l]{DL-based classification\\ (tabular data, face images)}  & Model & Interpretable white-box fairness testing through biased neuron identification & \cite{NeuronFairlink} \\
DeepInspect \cite{icseTianZOKR20}& \makecell[l]{DL-based image\\ classification}  & Model & Detecting class-based bias in image classification & \cite{DeepInspectlink} \\
CMA \cite{nipsVigGBQNSS20}& Language models  & Model & Detecting which parts of DNNs are responsible for unfairness & \cite{CMAlink}\\
FairNeuron \cite{icse22fair3} &\makecell[l]{DL-based classification\\ (tabular data)}   & Model & Detecting neurons and data instances responsible for bias & \cite{FairNeuronlink} \\
RULER \cite{guanhongfse22} & \makecell[l]{DL-based classification\\ (tabular data)}  &  Model & Test input generation by discriminating sensitive and non-sensitive attributes & \cite{rulerlink}\\
TestSGD \cite{tosemmengdi} & \makecell[l]{DL-based classification\\ (Tabular data, text)} &  Model & Interpretable testing of DNNs against subtle group discrimination & \cite{TestSGDlink}\\
DICE \cite{icse23Monjezi} & \makecell[l]{DL-based classification\\ (tabular data)} & Model & Information-theoretic fairness testing and debugging of DNNs & \cite{dicelink}\\
ASTRAEA \cite{9678017} & NLP systems & Model & Grammar-based discriminatory instance generation for NLP systems & \cite{ASTRAEAlink}\\
MT-NLP \cite{ijcai0004WL20} & NLP systems & Model &  Metamorphic testing of fairness violation in NLP systems& \cite{MTNLPlink} \\
BiasFinder \cite{9653830} & Sentiment analysis & Model & Metamorphic test generation to uncover bias of sentiment analysis systems & \cite{BiasFinderlink} \\
BiasRV \cite{sigsoftYangA021} & Sentiment analysis  & Model & Uncovering biased sentiment predictions at runtime & \cite{biasrvlink}\\
NERGenderBias \cite{mehrabi2020man} & Name entity recognition   & Model & Measuring gender bias in named entity recognition &\cite{NERGenderBiaslink}\\
CheckList \cite{aclRibeiroWGS20} & NLP systems & Model & Behavioral testing (including fairness testing) of NLP models & \cite{checklistlink} \\
DialogueFairness \cite{colingLiuDFLLT20} & Conversational AI  & Model & Testing gender and linguistic (racial) bias in dialogue systems & \cite{DialogueFairnesslink}\\
BiasAsker \cite{biasaskerfse23}  & Conversational AI  & Model &  Fairness testing of conversational AI systems &\cite{biasaskerlink}\\
REVISE \cite{eccvWangNR20} & CV datasets &  Data & Detecting object-, gender-, and geography-based bias in CV datasets & \cite{reviselink}\\
AequeVox \cite{faseRajanUC22} & Speech recognition & Model & Comparing the robustness of speech recognition systems for different groups & \cite{AequeVoxlink} \\
\hline
\end{tabular}
\end{table*}

\subsection{Open-source Testing Tools}\label{opentools}
There is a recent proliferation of open-source tools for supporting fairness testing. Nevertheless, Lee and Singh~\cite{chiLeeS21} demonstrated that there is a steep learning curve for practitioners to use these fairness tools.  Presently, there is a lack of guidance on tool adoption \cite{chiLeeS21}.

To address this gap, we provide a summary of 41 open-source fairness testing tools in this section, aiming to assist fairness researchers and practitioners in selecting the most suitable tools. The details of these tools are presented in Table \ref{tools_info}. The table includes fairness testing tools for various domains, including general ML (e.g., FairTest \cite{TramerAGHHHJL17}), DL (e.g., ADF \cite{icseZhangW0D0WDD20} and EIDIG \cite{isstaZhangZZ21}), natural language processing (e.g., ASTRAEA \cite{9678017} and BiasFinder \cite{9653830}), computer vision (e.g., REVISE \cite{eccvWangNR20}), and speech recognition (e.g., AequeVox \cite{faseRajanUC22}).

\section{Research Opportunities}\label{chall}
Fairness testing remains in a relatively embryonic state. Research in this area is experiencing rapid growth, so there are plenty of open research opportunities. In this section, we outline the challenges for fairness testing and present promising research directions and open problems.

\subsection{Absence of or with Multiple Sensitive Attributes}

\noindent \textbf{Fairness testing in the absence of sensitive attribute information.} Existing fairness testing techniques rely on the existence of sensitive attributes, but in practice, this information might be unavailable or imperfect for many reasons \cite{fatAwasthiBKM021}. On the one hand, the data may be collected in a setting where the sensitive attribute information is unnecessary, undesirable, or even illegal, considering the recently released regulations such as GDPR (General Data Protection Regulation) \cite{voigt2017eu} and CCPA (California Consumer Privacy Act) \cite{goldman2020introduction}. On the other hand, users may withhold or modify sensitive attribute information, for example, due to privacy concerns or other personal preferences.
To tackle this issue, a straightforward solution is to first use existing demographic information inference techniques (e.g., gender inference, race inference, and age inference) to infer the sensitive attribute and then apply fairness testing techniques. However, existing inference techniques may not be fully satisfactory, and their application scenarios remain limited \cite{wwwChenLALML18}. Moreover, building a model to infer sensitive information leaves open the possibility that the model may ultimately be used more broadly, with possibly unintended consequences \cite{fatAwasthiBKM021}. Therefore, more research is needed to tackle fairness testing in the absence of sensitive attribute information.

\noindent  \textbf{Fairness testing with multiple sensitive attributes.} Software systems can have multiple sensitive attributes that need to be considered at the same time \cite{intersectionalsurvey,corrabs230801923}.  Human attributes, such as sex, race, and class, intersect with one another, and unfair software systems built into society lead to systematic disadvantages along these intersecting attributes \cite{crenshaw2013demarginalizing,kimberly1989demarginalizing}. However, existing fairness testing work often tackles a single sensitive attribute at a time. To the best of our knowledge, there has been a little work that explores fairness testing for compounded or intersectional effects of multiple sensitive attributes \cite{cabrera2019fairvis,guanhongfse22,tosemmengdi}, leaving an interesting research opportunity for the community.  
Moreover, drawing parallels with historical legal discourse, the intersectionality of protected attributes reveals the challenges of avoiding over-segmentation when assessing fairness. In 1976, Judge Harris Wangelin expressed concerns about creating new protected groups \cite{judgecase}, specifically considering the implications of the `curse of dimensionality.' Much like the legal system's cautious approach to potential Pandora's boxes of new protected classes, fairness testing grapples with complexities in addressing compounded or intersectional effects of multiple sensitive attributes.

\subsection{Test Oracle for Fairness Testing}
Existing work mainly employs metamorphic relations as pseudo oracles or uses statistical measurements as indirect oracles of fairness testing, which both involve human ingenuity. It is an open challenge to design automatic techniques for constructing reliable oracles for fairness testing.

Furthermore, the emergence of manually-defined oracles for fairness testing brings a challenge for test oracle selection. For instance, the IBM AIF360 toolkit alone offers more than 70 fairness measurements \cite{AIF360link,corrbs181001943}, and the research community continues to introduce novel measurements. However, it is impractical to utilize all existing measurements as test oracles for fairness testing. Moreover, while each measurement may be suitable in a specific context, many of them cannot be simultaneously satisfied \cite{sigsoftBrunM18}. Hence, an important area for the research community is the development of automatic techniques for constructing reliable oracles in fairness testing. It is worth noting that one could argue that this challenge falls within the realm of requirements engineering rather than testing. However, in reality, fairness testing is frequently explored in natural settings, where upfront requirements engineering processes are not always assumed or followed strictly. Fairness testing involves investigating and evaluating fairness concerns in real-world systems or datasets, which may lack comprehensive and formal requirements. As a result, fairness testing needs to adapt to address the unique challenges that arise in these real-world contexts, where strict adherence to traditional requirements engineering may not be practical or feasible.

\subsection{Test Input Generation for Fairness Testing}
\noindent \textbf{Generation of natural inputs.} Despite the existence of various techniques for test input generation in fairness testing, there is no guarantee that the generated instances are legitimate and natural. Particularly, in Table \ref{testg_summary}, it is evident that most test input generation techniques are based on the causal fairness definition, which requires generating pairs of instances that differ solely in sensitive attributes. However, it remains uncertain whether altering only the sensitive attribute is sufficient to generate inputs that are truly natural.

Furthermore, existing techniques \cite{isstaZhangZZ21,9506918,icseZhangW0D0WDD20} primarily rely on perturbing input features, without explicitly constraining the magnitude of the perturbation. As long as the generated instances can induce the intended output behavior, such as flipping the predicted outcome after modifying sensitive attribute information, they are considered effective. However, this approach may overlook real-world constraints, potentially resulting in generated instances that do not align with reality (e.g., granting a loan to a 10-year-old individual).

As a result, open problems arise regarding how to generate test inputs for fairness testing that are both legitimate and natural. Researchers need to address the challenges of ensuring the generated instances adhere to real-world constraints while still accurately assessing fairness. Additionally, automating the evaluation of the naturalness of generated test inputs is another important area of exploration, enabling more reliable and efficient fairness testing methodologies.

\noindent \textbf{Exploration of more generation techniques.} As mentioned earlier, most test input generation techniques in fairness testing focus on the causal fairness definition. In contrast, test input generation for counterfactual fairness is relatively unmatured and more challenging. It requires researchers to conduct causal analysis of features and consider the causal relationships among them when altering sensitive attributes during generation. Moreover, when dealing with multiple sensitive attributes simultaneously, the task becomes even more difficult as changes in features need to account for the causal impacts from multiple sensitive attributes.

Furthermore, there is a research opportunity to design test inputs specifically for group fairness testing. While group fairness can be evaluated using real-world collected data, obtaining such data is not always easy. Given the limited work on generating test inputs for group fairness, there is still much potential for exploration in this area.



\subsection{Test Adequacy for Fairness Testing}
Test adequacy is a well-explored concept in traditional software testing, focusing on evaluating the coverage provided by existing tests  \cite{tseZhangHML22}. Adequacy criteria not only offer confidence in testing activities but also serve as a guide for test generation. However, in the domain of fairness testing, the issue of test adequacy remains an open problem, and to the best of our knowledge, no research has specifically addressed this area.

To address this challenge, one approach could be to adapt traditional software test adequacy metrics or ML test adequacy metrics for fairness testing. For instance, traditional software testing has proposed metrics such as line coverage, branch coverage, and dataflow coverage \cite{tse0050ZHH0019}, while ML testing has introduced metrics such as neuron coverage, layer coverage, and surprise adequacy for deep learning models \cite{tseZhangHML22}. Neuron coverage assesses the extent to which neurons in a deep learning model are exercised by a test suite, while layer coverage measures the coverage of different layers. Surprise adequacy, on the other hand, evaluates the coverage of discretized input surprise range for deep learning models \cite{tseZhangHML22}.

However, there is currently no empirical evidence to support the applicability and effectiveness of these metrics in assessing the ability to detect fairness bugs and the sufficiency of fairness testing. Further research is needed to investigate and validate the suitability of these metrics in the context of fairness testing. Additionally, novel metrics tailored specifically for fairness testing may need to be developed to capture the unique characteristics and requirements of assessing fairness in ML software.

\subsection{Test Cost Reduction} Test cost poses a significant challenge in fairness testing of ML software. The process of assessing fairness often entails retraining ML models, repeating the prediction process, or generating extensive data to explore the vast behavioral space of the models. However, thus far, there has been no research dedicated to reducing the cost of fairness testing. It would be intriguing to explore specific techniques for test selection, prioritization, and minimization that can effectively reduce the cost of fairness testing without compromising test effectiveness.

Moreover, as discussed in Section \ref{programtesting}, there is an escalating demand for deploying intelligent software systems on platforms with limited computing power and resources, such as mobile devices. Several studies \cite{corrabs210607849,abs201003058,corrabs220101709,corrabs210607849} have addressed fairness testing in such scenarios. This presents a fresh challenge for the research community: how to conduct fairness testing effectively on diverse end devices, including those with restricted computing power, limited memory size, and constrained energy capacity. Addressing this challenge requires innovative approaches and techniques that can adapt fairness testing methodologies to accommodate the limitations and constraints of these resource-constrained platforms.

\subsection{Fairness and Other Testing Properties}
\noindent \textbf{Testing fairness repair techniques with more properties considered.} After fairness repair techniques have been applied to software systems, fairness testing is often performed again. In this process, testers may also take  ML performance (e.g., accuracy) into consideration \cite{sigsoftHortZSH21,zhenpengmaat22}, because it is well-known that fairness improvement is often at the cost of ML performance \cite{sigsoftHortZSH21}. However, in addition to ML performance, there are also many other properties important for software systems, including robustness, security, efficiency, interpretability, and privacy \cite{tseZhangHML22}. The relationship between fairness and these properties is not well studied in the literature, and thus these relationships remain less well understood. Future research is needed to uncover the relationships and perform the testing with these properties considered. 
The determination of the properties to be considered needs the assistance of requirements engineers.

\noindent \textbf{Fairness and explainability.} Explainability is defined as that users can understand why a prediction is made by a software system \cite{gilpin2018explaining}. Like fairness, it has also been an important software property required by recent regulatory provisions \cite{MittelstadtRW19}. Because application scenarios that demand fairness often also require  explainability, it would be an interesting research direction to consider fairness and explainability together. Many existing fairness testing studies just generate discriminatory instances that reveal fairness bugs in the software under test, but do not explain why these instances are unfairly treated by the software. 

In this case, software engineers have relatively little guidance on the production of targeted fixes to repair the software. Improving the explainability behind the unfair software outcomes can help summarize the reasons for fairness bugs, produce insights for fairness repair, and help stakeholders without technical backgrounds (e.g., product managers, compliance officers, and policymakers) understand the software bias simply and quickly.

\subsection{Fairness Testing of More Applications}
The majority of existing fairness testing work has concentrated on tabular data, natural language processing systems, and computer vision systems. However, fairness, as a critical non-functional property, should be considered across a broader spectrum of software systems, including speech recognition systems, video analytic systems, multi-modal systems, and recommendation systems. Additionally, existing fairness testing studies have predominantly centered around classification tasks. However, fairness is a crucial concern that should be examined in various machine learning tasks, including regression and clustering as well as emerging cutting-edge AI technologies such as 
Large Language Models (LLMs). 

LLMs, extensively studied in academic literature and widely adopted in various applications, have recently raised concerns about fairness \cite{journalscorrabs230810149}. Given the significance of fairness in LLMs, OpenAI is actively seeking expertise to address these concerns and foster the development of fair LLMs~\cite{openaifairlink}. However, fairness testing for LLMs poses unique challenges. Firstly, LLMs are often open-domain systems that offer diverse functionalities. For instance, ChatGPT can engage in a variety of conversations with humans on a broad spectrum of topics. This characteristic poses a challenge when designing test oracles and input generation techniques for fairness testing, given the need to account for the extensive range of subjects and functionalities in LLMs. In contrast, existing fairness testing techniques are often tailored to specific tasks, potentially falling short in comprehensively evaluating the performance of LLMs across their diverse capabilities. Secondly, LLMs can generate a spectrum of responses, including those that may appear vague or unrelated, often influenced by pre-defined protection mechanisms regarding sensitive topics. This diversity in responses poses a challenge in automatically discerning whether the LLM output exhibits bias (i.e., the test oracle problem). While Wan et al. \cite{biasaskerfse23} have proposed solutions to address this challenge, their focus remains on tackling test oracle problems related to Yes-No questions, Choice-questions, and Wh-questions. Regrettably, this leaves other question types, such as prediction questions, explanatory questions, and recommendation questions, unexplored. Thirdly, the majority of large-scale LLMs are not open-sourced, leading to opacity in their underlying mechanisms and low explainability. This characteristic poses a challenge in designing fairness testing techniques, limiting practitioners to developing black-box approaches based solely on the observed responses of LLMs. Moreover, data testing and ML program testing are not applicable to such LLMs.

\subsection{More Fairness Testing Activities and Components}
The current body of research primarily focuses on offline fairness testing. There is a pressing need for more research in the realm of online fairness testing, as it can provide valuable insights to guide software maintenance and facilitate the evolution of software systems.

Furthermore, researchers have an opportunity to extend fairness testing investigations to include additional testing activities that have been extensively studied in traditional software testing but rarely explored in the context of fairness testing. For instance, exploring bug report analysis \cite{zhang2015survey}, bug triage \cite{jeong2009improving}, and test evaluation \cite{tseZhangHML22} in fairness testing can contribute valuable insights to the field.

Additionally, there exists a research gap in the fairness testing components. While testing of ML frameworks has received considerable attention in traditional ML testing \cite{icsePhamLQT19,sigsoftWangYCLZ20,NejadgholiY19,icseWangLQP022}, its application in fairness testing remains underexplored. Furthermore, exploring non-ML component testing within the context of fairness testing presents a promising research direction that warrants further investigation.


\subsection{More Fairness Testing Tools}
Existing fairness testing tools (listed in Table \ref{tools_info}) tend to require programming skills, and thus are unfriendly to non-technical stakeholders.
However, fairness testing research includes many non-programmer stakeholders and contributors such as compliance officers, policymakers, and legal practitioners.

\section{Discussion}
\subsection{Stakeholders in Fairness Testing}
Fairness testing goes beyond the pure view of test engineers, involving a range of stakeholders~\cite{SarroRE}. Given that unfairness can stem from data or algorithms, data scientists and algorithm designers can play pivotal roles in detecting potential biases and providing valuable insights for fairness testing. Additionally, legal practitioners, compliance officers, and policymakers can be queried as crucial stakeholders, ensuring that fairness aligns with encoded laws, regulations, and policies.

Moreover, it is imperative to expand our focus beyond conventional algorithmic testing. ML software users, directly impacted by algorithmic decisions, offer unique insights and real-world experiences that extend beyond the scope of algorithmic scrutiny alone. This is exemplified by the book `Algorithms of Oppression' \cite{noble2018algorithms}, which provides a good example illustrating that, despite potential testing of its algorithms, Yelp's review system demonstrated unfairness in its waiting system concerning a specific user population, adversely affecting a local hairdresser store. This emphasizes the significance of community perspectives in revealing biases not immediately evident through algorithmic testing alone. Incorporating user-centered design principles and establishing feedback loops during development and testing are crucial for a more comprehensive approach to fairness testing.

\subsection{Algorithmic Fairness and Societal Fairness}
Certain researchers posit that algorithms inherently reflect the biases embedded in their social context \cite{lewis1977forces}. Consequently, they argue that addressing algorithmic bias holds limited value unless we first address and rectify the societal issues that influence the selection and deployment of these algorithms. However, we believe that addressing algorithmic fairness is not misguided but complementary to tackling societal fairness. While algorithms may be inherently biased due to societal bias, addressing algorithmic fairness focuses on short-term solutions within technology to reduce discrimination. Recognizing that societal fairness is a complex, long-term endeavor, both efforts—algorithmic and societal fairness—should coexist for a fairer future.

In addition, algorithms are technologies created by people. Even if algorithms reflect societal biases, it does not absolve the creators of algorithms from their responsibilities. Algorithmic fairness research emphasizes holding them accountable for addressing bias in algorithms.

Furthermore, the algorithmic view of fairness raises awareness about societal consequences in decision-making. It contributes to a comprehensive effort for fairness, acknowledging that improving algorithms is a tangible, achievable step towards broader societal change.

We also recognize that addressing algorithmic fairness fundamentally requires an improvement in societal fairness, emphasizing crucial social factors such as power structures and social justice. The technical dimensions of fairness testing necessitate a holistic consideration of the societal contexts in which these algorithms function.

A significant aspect involves acknowledging that AI technologies often stem from the directives of individuals or entities in positions of power \cite{sloane2022make}. To rectify this power imbalance, a profound shift is imperative. It extends beyond merely consulting those affected by AI at the outset; active participation in the decision-making process is essential. This entails empowering individuals to select the problems addressed and guide the entire developmental trajectory.

Furthermore, the involvement of the most relevant participants becomes paramount to ensure that AI training data authentically mirrors the diversity of perspectives in society \cite{sloane2022make}. This principle underscores the significance of integrating lived experiences into the development of AI systems. This approach not only mitigates the risk of participation-washing but also fosters a more inclusive methodology for fairness testing.

\subsection{Threats to Validity}
As presented in Section \ref{methodology}, the methodology of this paper includes two primary phases: paper collection and paper analysis. In this section, we discuss potential threats associated with these two phases.

\noindent \textbf{Paper collection:} Our collection process involves the selection of research papers from the widely-used DBLP database, which encompasses arXiv and over 1,800 journals, as well as 5,800 academic conferences and workshops in the field of Computer Science. However, it is important to acknowledge that this approach may overlook papers published in other venues. Additionally, there is the possibility that our manually defined search strings may not encompass all the relevant studies within our research scope. To address these potential threats, we employ both backward snowballing and forward snowballing to further identify transitively dependent papers. Moreover, we proactively contact the authors of our collected papers to solicit additional papers that fall within the scope of our survey.

\noindent \textbf{Paper analysis:} The process of manually analyzing research papers to extract relevant information to be included in the survey is potentially open to human bias. To mitigate this risk, each paper undergoes analysis by two separate authors, and any disagreement that arises during the analysis is resolved through discussions involving other co-authors, all of whom have previously published fairness-related papers in top-tier SE venues. Furthermore, all authors independently conduct a thorough review of the survey's content to identify and rectify any potential issues.  We also share our survey with the authors of the collected papers to ensure that our descriptions of their work are accurate.

\section{Conclusion}
We have presented a comprehensive survey of \num papers on fairness testing. We summarized the current research status in the fairness testing workflow (including test input generation and test oracle identification) and testing components (including data testing, ML program testing, and model testing). We analyzed trends and promising research directions for fairness testing. We also listed public datasets and open-source tools that can be accessed by researchers and practitioners interested in this topic. We hope this survey will help researchers from various research communities become familiar with the current status and open opportunities of fairness testing.  

\section*{Acknowledgment}
We shared our work with the authors of the papers we surveyed in order to check for accuracy and omission, and we would like to thank those authors who kindly provided comments and feedback on earlier drafts of this paper.
Zhenpeng Chen, Federica Sarro, and Mark Harman are supported by the ERC Advanced Grant No.741278 (EPIC: Evolutionary Program Improvement Collaborators). Max Hort is supported by the Research Council of Norway through the secureIT project (IKTPLUSS \#288787).

\bibliographystyle{ACM-Reference-Format}
\bibliography{fairnesssurvey}

\end{document}